\documentclass[12pt]{article}

\usepackage[english]{babel}
\usepackage[utf8x]{inputenc}
\usepackage{amsmath}
\usepackage{graphicx}
\usepackage[colorinlistoftodos]{todonotes}
\usepackage[margin=1in]{geometry}
\usepackage{setspace}
\usepackage{indentfirst}
\usepackage{listings}
\usepackage{amsfonts}
\usepackage{bbold}
\usepackage{pdflscape}  
\usepackage{longtable}  
\usepackage{array}
\usepackage{tabularx}
\usepackage[colorlinks=true, linkcolor=blue, urlcolor=blue]{hyperref}
\usepackage{float}
\usepackage{subcaption}

\title{Multimarket Contact, Merger, and Airline Collusion
\thanks{I cannot express enough thanks to Prof. Daniel Waldinger, my dissertation tutor, and Prof. Viplav Saini, the instructor of the senior honors tutorial, for their sincere, precise, and inspirational advice on my dissertation. Their guidance has helped me stay on the right path throughout its completion. Additionally, I offer my sincere gratitude to Prof. Elena Manresa, who provided invaluable advice on the empirical strategies in this dissertation.

I want to thank my amazing colleagues in the economics honors seminar, especially Aaron, Eric, Future, Michael, and Sue, whose passionate and accommodating conversations during the seminar sparked countless ideas for me. My sincere appreciation also goes to Zhijun "Walden" He, who was the first reader of this dissertation and never hesitated to share his thoughts with me.

Moreover, I would like to thank my parents for their generous financial and intellectual support. The unyielding encouragement from Heidi, my sister, will never be forgotten. Last but not least, I want to express my gratitude to Weilai Zhao, my life partner. Her heart shines like the sun, providing me with the most intimate and timely help and guidance over these years.}}
\author{Ziyu "Harry" Yan}
\date{May 14, 2024}

\begin{document}

\newcolumntype{Y}{>{\centering\arraybackslash}X}  

\onehalfspacing
\maketitle

\begin{abstract}
This thesis investigates the dynamics of multimarket contact and airline mergers on collusive pricing of airlines. In align with Bernheim and Whinston (1990) and Athey et.al. (2004), it detects collusive pricing via pairwise price difference and price rigidity.

The piece of work extends previous work by incorporating additional controls such as distinction between non-stop and stopover itineraries and detailed market concentration measures. The findings confirm a significant relationship between multimarket contact and reduced price differences, indicating collusive equilibria facilitated by frequent interactions across markets. Moreover, the results highlight that airlines exhibit more collusive behavior when pricing non-stop flights, and are more likely to attain tacit collusion when they approaches duopoly in a particular market.

The study also explores the effects of airline mergers on collusion, employing an event study methodology with a difference-in-difference (DID) design. It finds no direct evidence that mergers lead to increased collusion among unmerged carriers. However, it reveals that during and after the merger process, carrier pairs between merged and unmerged carriers are more likely to collude compared to pairs of unmerged carriers.     
\end{abstract}

\newpage

\section{Introduction}

Collusive behavior of firms is one of the central issues of industrial organization. The airline industry is an ideal case to study collusion given its scope, transparency, public attention, and level of competition. While collusion per se is prohibited under the Section 1 of the Sherman Act, tacit collusion is instead evaluated under the rule of reason. Literature has effectively attributed the collusion of airlines to their multimarket contacts \href{https://jonwms.web.unc.edu/wp-content/uploads/sites/10989/2021/06/CollusivePatterns_IJIO.pdf}{\textcolor{blue}{Ciliberto et. al. (2019)}}. Furthermore, entering the 21st century, several mergers take place among some well-known U.S. carriers, arousing debates on antitrust review regarding the effect of these mergers. Therefore, through a two-prolonged approach, this paper aims to provide a more robust and updated analysis on multimarket contact and collusive pricing, as well as extending the discussion to relationship between airlines merger, and collusive pricing.

Therefore, by analyzing ticket price variations and multimarket contact. the first part of this paper extends \href{https://jonwms.web.unc.edu/wp-content/uploads/sites/10989/2021/06/CollusivePatterns_IJIO.pdf}{\textcolor{blue}{Ciliberto et. al. (2019)}}'s work. The model detects collusion by observing price difference and price rigidity of differentiated goods. The first improvement is the distinctions between non-stop and stopover markets over a particular city pair. This is because consumers choose to take connecting flights may behave differently comparing with consumers choose to take a non-stop flight. For instance, they may be more sensitive to ticket prices. Furthermore, usually, there are more carrier options when taking connecting flights comparing with taking non-stop flights, which may influence the level of collusion. The second improvement is a more detailed measurement on market concentration. For multiple economic incentives such as higher market power and lower marginal cost provided by economy of scale, a firm with dominant share in a market is likely to behave different from firms with little share in the same market. Therefore, this paper introduces the total share and the relative share of a particular pair of carriers to replace the Hirschmann-Herfindahl Index (HHI) as controlling variables on market concentration. 

The second part of the study assesses the impact of airline mergers on collusion using event study methodology with a difference-in-difference (DID) design. This part continues the empirical strategy to identify collusion by pairwise price difference between two carriers in a year-quarter-market-carrier pair level observation. The time dummies in DID helps capturing competitive dynamics during the regulatory approval process, operation integration, and post-merger. The treatment dummies distinguishes the carrier pair between two merged carriers, merged and unmerged carrier, and two unmerged carriers, which 
addresses asymmetric effect of the merger on the collusive patterns between different types of carriers.

\section{Literature Review}

The research aims to contribute to the intersection of two areas. First is the empirical analysis on collusive behavior in the airline industry, and second is the impact of airlines merger and acquisition on their collusive behavior.

In the first literature, there are more papers attempting to discover a structural pattern for the airlines’ collusive behavior. For example, \href{https://doi.org/10.1016/0167-7187(93)90017-7}{\textcolor{blue}{Brander and Zhang (1990)}}
, by adapting conduct parameters to analyze the cross-section data from the 1985 U.S. airline industry, discovers that the pricing of airlines is closer to Cournot than either Bertland, cartel or competitive model. Furthermore, \href{https://doi.org/10.2307/2555469}{\textcolor{blue}{Brander and Zhang (1993)}} extended the discussion into a dynamic setting and adapted a regime-setting (RS) model developed by \href{https://doi.org/10.2307/1911462}{\textcolor{blue}{Green and Porter (1984)}} to measure the level of collusion between airlines. In the RS model, firms shift from the collusive phase to a punishing phase if the price falls below some trigger level. The paper confirms that the pricing behavior of airlines is closer to Cournot model. It also detects the periodic collusive pattern of airlines, as they revert to Cournot behavior (which implies a punishing phase) in some periods. However, both papers above only discussed symmetric duopoly cases, and treated the elasticity and the cost-tapering factor as constant over route and time. Another issue of \href{https://doi.org/10.2307/2555469}{\textcolor{blue}{Brander and Zhang (1993)}} is, as \href{http://www.jstor.org/stable/3700628}{\textcolor{blue}{Athey et. al. (2004) }} discovered that price rigidity is an important identifier of collusive pricing behavior, identifying the punishing phase in RS-model merely by absolute price level has become questionable. 
 
Based on the idea from \href{https://doi.org/10.2307/2555490}{\textcolor{blue}{Bernheim and Whinston (1990)}}, \href{http://www.jstor.org/stable/43186481}{\textcolor{blue}{Ciliberto and Williams (2014) }} provides a structural analysis of multimarket contact and collusion in the airline industry. The model assumes that firms price as Bertland-Nash competitors, then it predicts a significant correlation between multimarket contact and collusive pricing of airlines.When airlines have little multimarket contact, they do not cooperate in setting fares. However, when airlines have a significant amount of multimarket contact, they can sustain near-perfect cooperation in setting fares. After that, the paper conducts an empirical analysis of both prices and demand in a particular market. The result verifies the structural model they have proposed above, and further reveals that while legacy carriers tend to set prices cooperatively with each other, low-cost carriers (LCC) seldom set prices cooperatively with legacy carriers. Based on this observation, \href{http://www.jstor.org/stable/43186481}{\textcolor{blue}{Ciliberto and Williams (2014) }} also suggest that from their structural model, a merger analysis could be extended. They noticed that an increase in multimarket between legacy carriers resulted in almost no change in fares, whereas the same change in multi-market contact LCC and legacy carriers resulted in large increases in fares. 

Of course, there are also a few empirical papers related to the collusive behavior of airlines. \href{https://doi.org/10.2307/2118466}{\textcolor{blue}{Evans and Kessides (1994)}} discusses the impact of multimarket contact on the price of air tickets. However, it limits its regression to explain the price, and did not find significant evidence of collusive behavior of airlines. Recent progress is made by Ciliberto et. al. (2019). By identifying the collusive behavior of airlines by their price difference as well as price rigidity, the paper correlates collusion with  multimarket contact, code-sharing agreement, and fixed effects. Compared with the empirical part of \href{http://www.jstor.org/stable/43186481}{\textcolor{blue}{Ciliberto and Williams (2014) }}, this paper sketches collusive behavior with pairwise price difference and price rigidity, while earlier research such as \href{https://doi.org/10.2307/2555469}{\textcolor{blue}{Brander and Zhang (1993)}}  identifies collusion by comparing the price and quantity demanded in the market with a counterfactual they have constructed. However, a limitation of this paper is, it did not distinguish non-stop markets from connecting markets when measuring multimarket contacts, while most consumers have a preference for non-stop flights under the same price. Moreover, \href{https://jonwms.web.unc.edu/wp-content/uploads/sites/10989/2021/06/CollusivePatterns_IJIO.pdf}{\textcolor{blue}{Ciliberto et. al. (2019)}} believe that the impact of airline mergers could be explained by an increase of multimarket contact. However, the impact of mergers is more profound than the change in multimarket contact, and this intended research aims to assess it comprehensively.

In the second literature, generally, there are more papers about the merger of airlines from a multitude of perspectives, such as economy of scale as incentive of merger \href{https://doi.org/10.1016/j.tre.2012.02.002}{\textcolor{blue}{(Merkert and Morrell, 2012)}}, productivity and efficiency \href{https://doi.org/10.1016/j.jairtraman.2022.102226}{\textcolor{blue}{(Khezrimotlagh et.al., 2022)}}, consumer welfare related to flight frequency \href{https://doi.org/10.1016/S0167-7187(03)00002-X}{\textcolor{blue}{(Richard, 2003)}}, and network structure \href{https://doi.org/10.1016/S0167-7187(03)00002-X}{\textcolor{blue}{(Shaw and Ivy, 1994)}}. Some of the empirical literature addresses the impact of mergers on prices and market power. However, few of them focus directly on the collusive behavior of airlines. For example, \href{http://www.jstor.org/stable/2117533}{\textcolor{blue}{Kim and Singal (1993)}} examines price changes associated with airline mergers, and find that price increased on routes served by the merging firms relative to a control group of routes unaffected by the merger. The paper attributes such change to the increasing market power of the firm. The impact of increasing market power of the carrier offsets the gain in consumer welfare resulting from the efficiency increase after the merger, making consumers worse off. 
 
Another notable literature is \href{https://doi.org/10.1007/s11151-013-9380-1}{\textcolor{blue}{Luo (2014)}}, which focuses specifically on the influence of the Delta/Northwest merger in 2008 on the entire aviation market. The paper reveals that the merger only caused a small increase in the fares of non-stop flights, which may be because two carriers have little overlapping services before their mergers. \href{https://doi.org/10.1007/s11151-013-9380-1}{\textcolor{blue}{Luo (2014)}} further suggests that the effect on ticket prices will be more prominent in the fare of connecting flights, in which two carriers have more overlapping services. The paper also points out that the (small) positive price effect in non-stop flights may be caused by the exit of LCC as an effect of merger. Notably, if we see the price increase as a signal of collusive pricing behavior, the reasoning provided by \href{https://doi.org/10.1007/s11151-013-9380-1}{\textcolor{blue}{Luo (2014)}} is contrary to the reasoning by \href{https://jonwms.web.unc.edu/wp-content/uploads/sites/10989/2021/06/CollusivePatterns_IJIO.pdf}{\textcolor{blue}{Ciliberto et. al. (2019)}}, because with little overlapping services, the merged carrier is ought to have considerably more multimarket contacts than before, which tends to increase the collusive behavior in the prediction by \href{https://jonwms.web.unc.edu/wp-content/uploads/sites/10989/2021/06/CollusivePatterns_IJIO.pdf}{\textcolor{blue}{Ciliberto et. al. (2019)}}. 

There are also papers that provide a structural analysis and simulation on the airlines’ merger. \href{https://sticerd.lse.ac.uk/seminarpapers/ei24052010.pdf}{\textcolor{blue}{Benkard et. al (2010)}} proposes an airline entry model to evaluate the long-run dynamic effects of airline mergers, and use it to simulate the effect of three proposed airline mergers in 2010. The paper asserts that a merger between two major hub carriers leads to increased entry by the other hub carriers, and can lead to substantial increased entry by low cost carriers, both effects offsetting some of the initial concentrating effects of the merger.

\section{Data}

The data utilized in the research comes from the Airline Origin and Destination Survey (DB1B) database of the U.S. Department of Transportation (USDOT) starting from 1993\footnote{Earlier data from DB1A starting from 1979 is not adapted because it does not distinguish ticketing carriers, which is the airline that a passenger purchases his/her ticket from, from the operating carriers in the record. This paper focuses on the pricing pattern of airlines, thus each itinerary should be attributed to its ticketing carrier. The emergence of charter airlines such as SkyWest which focus on operation and aircraft maintenance for flights that are scheduled, marketed and sold by a partner legacy carrier further justifies such a setup. In the records for these flights, the legacy carriers will be labeled as ticketing carriers, and charter airlines will be labeled as operating carriers. }. It is a $10\%$ sample of domestic ticket itineraries in each quarter containing origin, destination, and other details. In this paper, ticket-level DB1B data from the first quarter of 1993 to the second quarter of 2023 is first aggregated into time(year-quarter)- market-carrier level. For weighted average market fare in each aggregated observation, an average weighted by passengers in each ticket itinerary is used since some itineraries have more than one passenger. Fares are deflated by the consumer price index to be in 2017 U.S. dollars. Following earlier studies such as \href{https://www.jstor.org/stable/2951571?origin=crossref}{\textcolor{blue}{Berry(1992)}} and \href{https://jonwms.web.unc.edu/wp-content/uploads/sites/10989/2021/06/CollusivePatterns_IJIO.pdf}{\textcolor{blue}{Ciliberto et.al.(2019)}}, itineraries with exceedingly high (greater than $\$2,500$) and low (less than $\$25$) fares are removed in the aggregation process. Airlines and markets that do not represent a competitive presence in this industry are also excluded. Only 20 Largest airlines by average number of passengers transported per quarters is included\footnote{They are Southwest (WN), Delta (DL), American Airlines (AA), United (UA), U.S. Airways (US), Continental Airlines (CO), Alaska Airlines (AS), Northwest Airlines (NW), Jetblue (B6), Spirit Airlines (NK), Frontier Airlines (F9), AirTran Airlines (FL), America West Airlines (HP), Hawaiian Airlines (HA), Allegiant Air (G4), Trans World Airways (TW), American Trans Air (TZ), Aloha Airlines (AQ), Virgin America (VX), and Sun Country Airlines (SY).}.
These airlines above have contributed $98.72\%$ of all ticket-level itineraries. Furthermore, time-market-carrier-level observations with less than 30 passengers surveyed during a quarter are also excluded.

In this paper, to introduce a distinction between non-stop and stopover flights between the same city pair,  different from \href{https://jonwms.web.unc.edu/wp-content/uploads/sites/10989/2021/06/CollusivePatterns_IJIO.pdf}{\textcolor{blue}{Ciliberto et.al.(2019)}}, “market” is defined as a one-way trip between two airports with a particular flight segment. Therefore, it is “year-quarter-carrier- market-coupon” in the context of \href{https://jonwms.web.unc.edu/wp-content/uploads/sites/10989/2021/06/CollusivePatterns_IJIO.pdf}{\textcolor{blue}{Ciliberto et.al.(2019)}}. Such a change in the definition of markets resulted from multiple factors, such as expenditure structure amd more price-sensitive consumers, may motivate carriers to have different pricing strategies and hence collusive patterns regarding non-stop and stopover tickets. Figure 1 below has plotted the number of markets and nonstop markets over the time. In the second quarter of 2023, there are 17,242 markets, and 6,274 of them are non-stop markets. From the figure, a significant seasonal fluctuation in the number of markets can be observed, which implies a necessity of a fixed-effect design considering seasonality. 

\begin{center}
\includegraphics[width=\textwidth]{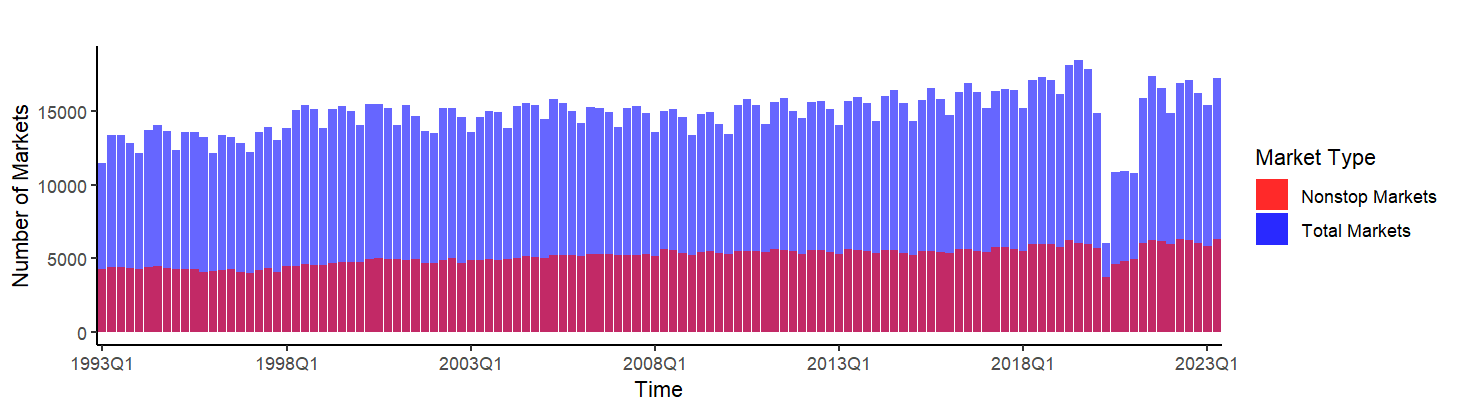}
{\small \textbf{Figure 1}: Number of Markets and Non-Stop Markets Over the Time}
\end{center}

On the other hand, when defining multimarket contact, this paper aligns with Evans and Kessides (1994) as well as \href{https://jonwms.web.unc.edu/wp-content/uploads/sites/10989/2021/06/CollusivePatterns_IJIO.pdf}{\textcolor{blue}{Ciliberto et.al.(2019)}}, defining multimarket contact as the number of unidirectional city pairs (regardless of stopovers) that two distinct carriers, $h$ and $k$, concomitantly serve at time $t$ and denoting as $MMC^{EK}_{hk,t}$. Such design is because stopover flights provided by one carrier is a substitution for non-stop flights provided by another carrier, which facilitates a market contact. Different pricing strategies regarding non-stop and stopover flights do not conflict with the market contact established between two carriers. Table 1 Below has displayed the (symmetric) matrix of the multimarket contacts for the 11 carriers active in the sample for the second quarter of 2023. For example, American and United concomitantly served 3,651 markets during the second quarter of 2008, then $MMC^{EK}_{AA, UA}$ equals to 3,651. Noticeably, $MMC^{EK}_{hk,t}$ is symmetric, which implies $MMC^{EK}_{hk,t}=MMC^{EK}_{kh,t}$.
 
For the robustness of the analysis, alternative definitions of multimarket contacts are also calculated. One alternative measurement is provided by dividing $MMC^{EK}_{hk,t}$ by the total number of markets served by firm $h$, and is denoted as $MMC^{CW}_{h \rightarrow k,t}$. Compared to $MMC^{EK}_{hk,t}$, it is asymmetric for any pair of airlines that each has a different network size. Another alternatives is to weight $MMC^{CW}_{h \rightarrow k,t}$ according to number of passengers or total revenue, which better reflects each market’s relative importance to the carrier in terms of passenger volume and revenue. Table 2 provides the complete carrier-pair $MMC^{CW}_{h \rightarrow k, t}$ in the second quarter of 2023. 

\begin{table}[ht]
\centering
\small
\caption{Pairwise Number of Multimarket Contacts $MMC^{EK}_{hk}$ in 2023Q2}
\vspace{12pt}
\begin{tabularx}{\textwidth}{ Y Y Y Y Y Y Y Y Y Y Y Y }
\hline
 & AA & AS & B6 & DL & F9 & G4 & HA & NK & SY & UA & WN \\
\hline
 \textbf{AA}  & 10317 &  &  &  &  &  &  &  &  &  &  \\
 \textbf{AS}  & 424 & 1387 &  &  &  &  &  &  &  &  &  \\
 \textbf{B6}  & 432 & 39 & 648 &  &  &  &  &  &  &  &  \\
 \textbf{DL}  & 5677 & 726 & 441 & 9311 &  &  &  &  &  &  &  \\
 \textbf{F9}  & 628 & 85 & 93 & 613 & 964 &  &  &  &  &  &  \\
 \textbf{G4}  & 30 & 17 & 6 & 29 & 22 & 1065 &  &  &  &  &  \\
 \textbf{HA}  & 43 & 41 & 4 & 59 & 0 & 0 & 177 &  &  &  &  \\
 \textbf{NK}  & 687 & 93 & 126 & 693 & 351 & 26 & 0 & 1058 &  &  &  \\
 \textbf{SY}  & 24 & 6 & 2 & 128 & 9 & 11 & 0 & 12 & 162 &  &  \\
 \textbf{UA}  & 3651 & 519 & 247 & 3272 & 612 & 26 & 60 & 566 & 16 & 5913 &  \\
 \textbf{WN}  & 4021 & 446 & 240 & 3714 & 733 & 88 & 114 & 744 & 20 & 2822 & 6408 \\
\hline
\multicolumn{12}{l}{\textit{Note: The off-diagonal numbers represents the number of markets served concomitantly by the}} \\
\multicolumn{12}{l}{\textit{carrier in the row and the carrier in the column. The numbers onthe diagonal are the total}} \\
\multicolumn{12}{l}{\textit{number of markets, including monopolistic markets, served by a carrier.}} \\
\end{tabularx}
\end{table}

\begin{table}[ht]
\centering
\small
\caption{Pairwise Number of Multimarket Contacts $MMC^{CW}_{h \rightarrow k}$ in 2023Q2}
\vspace{12pt}
\begin{tabularx}{\textwidth}{ Y Y Y Y Y Y Y Y Y Y Y Y }
\hline
 & AA & AS & B6 & DL & F9 & G4 & HA & NK & SY & UA & WN \\
\hline
 \textbf{AA}  & 1 & 0.041 & 0.042 & 0.550 & 0.061 & 0.003 & 0.004 & 0.067 & 0.002 & 0.290 & 0.317 \\
 \textbf{AS}  & 0.306 & 1 & 0.028 & 0.523 & 0.061 & 0.012 & 0.029 & 0.067 & 0.004 & 0.374 & 0.322 \\
 \textbf{B6}  & 0.667 & 0.060 & 1 & 0.681 & 0.144 & 0.009 & 0.006 & 0.194 & 0.003 & 0.381 & 0.370 \\
 \textbf{DL}  & 0.610 & 0.078 & 0.047 & 1 & 0.066 & 0.003 & 0.006 & 0.074 & 0.014 & 0.351 & 0.399 \\
 \textbf{F9}  & 0.651 & 0.088 & 0.096 & 0.636 & 1 & 0.023 & 0 & 0.364 & 0.009 & 0.635 & 0.760 \\
 \textbf{G4}  & 0.028 & 0.016 & 0.006 & 0.027 & 0.021 & 1 & 0 & 0.024 & 0.010 & 0.024 & 0.083 \\
 \textbf{HA}  & 0.243 & 0.232 & 0.023 & 0.333 & 0 & 0 & 1 & 0 & 0 & 0.339 & 0.644 \\
 \textbf{NK}  & 0.649 & 0.088 & 0.119 & 0.655 & 0.332 & 0.025 & 0 & 1 & 0.011 & 0.535 & 0.699 \\
 \textbf{SY}  & 0.148 & 0.037 & 0.012 & 0.790 & 0.056 & 0.068 & 0 & 0.074 & 1 & 0.099 & 0.123 \\
 \textbf{UA}  & 0.617 & 0.088 & 0.042 & 0.553 & 0.104 & 0.004 & 0.010 & 0.096 & 0.003 & 1 & 0.477 \\
 \textbf{WN}  & 0.627 & 0.070 & 0.037 & 0.580 & 0.114 & 0.014 & 0.018 & 0.115 & 0.003 & 0.440 & 1 \\
\hline
\multicolumn{12}{l}{\textit{Note: The $h-th$ row and $k-th$ column of the table represents $MMC^{CW}_{h \rightarrow k,t}$, which equals to }} \\
\multicolumn{12}{l}{\textit{$MMC^{EK}_{hk,t}$ divided by $h$.}} \\

\end{tabularx}
\end{table}

\begin{center}
\includegraphics[width=0.49\textwidth]{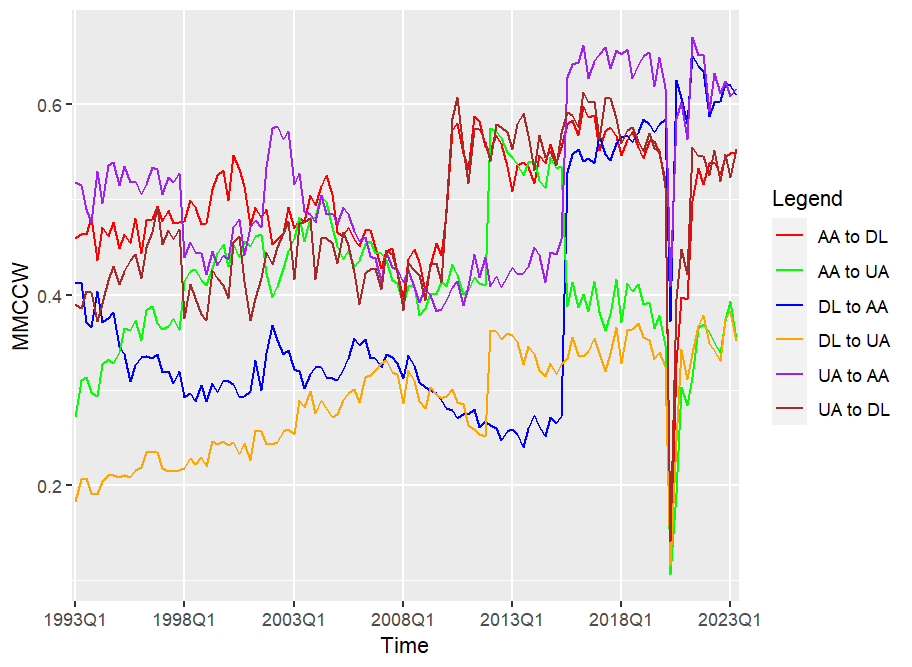}
\includegraphics[width=0.49\textwidth]{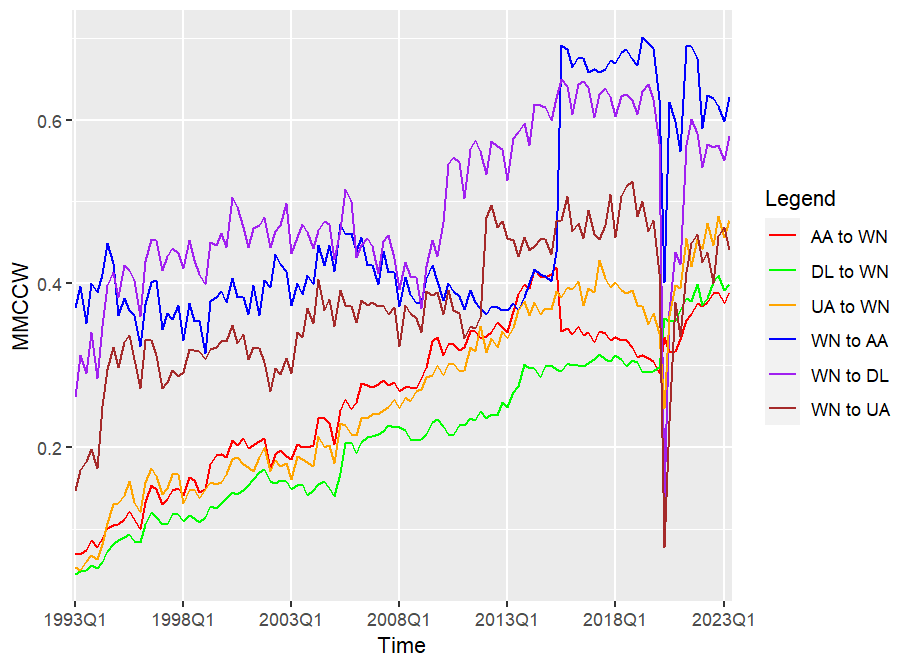}

{\small \textbf{Figure 2-3}: $MMC^{CW}$ Among Legacy Carriers and Between Legacies and Southwest} 
\end{center}

Figure 2 displays the $MMC^{CW}$ among American, Delta, and United, which are three largest legacy carriers in the U.S. The time series variation of them suggests multiple underlying influences on the multimarket contact, such as network expansion, seasonality, and merger. Figure 3, on the other hand, displays $MMC^{CW}$ between legacy carriers and Southwest, a rapidly growing Low-Cost Carrier.The figure exhibits that  $MMC^{CW}_{AA \rightarrow SW}$, $MMC^{CW}_{DL \rightarrow SW}$, and $MMC^{CW}_{UA \rightarrow SW}$ all follows a steady growing trend, and has reached the level of $MMC^{CW}_{SW \rightarrow AA}$ $MMC^{CW}_{SW \rightarrow DL}$, and $MMC^{CW}_{SW \rightarrow UA}$ after decades of expansion and acquisition. Relatively, $MMC^{CW}_{SW \rightarrow AA}$ $MMC^{CW}_{SW \rightarrow DL}$, and $MMC^{CW}_{SW \rightarrow UA}$ are more stable, implying that competition from legacy carriers has a constant meaningful presence across Southwest's network over the time.

\begin{table}[ht]
\centering
\small
\caption{Price Differences and Potential Facilitators of Collusion}
\vspace{12pt}
\begin{tabularx}{\textwidth}{l Y Y Y Y Y }
\hline
  & $\#$ Obs & Mean & Std. Dev. & Min & Max \\
\hline
\multicolumn{6}{l}{\textit{Year-Quarter-Market Carrier Sample (Unmatched)}}\\
 Mean Price & 4,131,161 & 237.981 & 84.425 & 25.004 & 2496.014 \\
 Carrier Passengers & 4,131,161 & 277.355 & 743.300 & 30 & 22101 \\
 Non-Stop Distance & 4,131,161 & 1285.969 & 776.73 & 9 & 8061 \\
  &  &  &  &  &  \\
\multicolumn{6}{l}{\textit{Pairwise Price Difference Sample}}\\
 $\Delta p$ & 3,877,723 & 43.144 & 44.448 & 0 & 933.742 \\
 $\mathbb{1}^{non-stop}$ & 3,877,723 & 0.106 & 0.308 & 0 & 1 \\
 $MMC^{EK}$ & 3,877,723 & 1591.755 & 1255.906 & 1 & 6097 \\
 $MMC^{CW}$ & 3,877,723 & 0.249 & 0.122 & 0.00004 & 0.921 \\
 Market Passengers & 3,877,723 & 808.4419 & 1222.062 & 60 & 80114 \\
 Relative Share & 3,877,723 & 0.540 & 0.254 & 0.002 & 1 \\
 Combined Share & 3,877,723 & 0.505 & 0.266 & 0.002 & 1 \\
 &  &  &  &  &  \\
\multicolumn{6}{l}{\textit{Pairwise Price Rigidity (Coefficient of Variation) Sample}}\\
 $CV$ & 217,886 & 0.205 & 0.115 & $7.53 \times 10^{-7}$ & 2.304422 \\
 $\mathbb{1}^{non-stop}$ & 217,886 & 0.160 & 0.367 & 0 & 1 \\
 $MMC^{EK}$ & 217,886 & 1677.266 & 1700.703 & 2 & 5677 \\
 $MMC^{CW}$ & 217,886 & 0.253 & 0.189 & 0.002 & 0.790 \\
 Market Passengers & 217,886 & 9066.55 & 38234.6 & 60 & 1954002 \\
 Relative Share & 217,886 & 0.564 & 0.266 & 0.003 & 1 \\
 Combined Share & 217,886 & 0.496 & 0.228 & 0.00003 & 1 \\
\hline
\multicolumn{6}{l}{\textit{Statistics for price differences use a year-quarter-market-carrier pair level of observation, while}}\\
\multicolumn{6}{l}{\textit{those for price rigidity use a market-carrier pair unit of observation. The multimarket contact}}\\
\multicolumn{6}{l}{\textit{statistics for price rigidity regression comes from 2023Q2, and the market passenger statistic of}}\\
\multicolumn{6}{l}{\textit{it comes from summing up the passenger volume for all times.}}
\end{tabularx}
\end{table}

The first panel of table 3 has summarized some statistics of the year-quarter-market-carrier level observations. In alignment with Werden and Froeb (1994), this paper first uses absolute difference in the price for each pair of carriers $hk$ in market $m$ at time $t$ as the dependent variable to narrate the collusive behavior of airlines, which is denoted as $\Delta p_{hk,mt}$ equals to $\|p_{h,mt}-p_{k,mt}\|$. The average of such pairwise absolute difference is around $\$43.14$. Then, as Athey et.al (2004) has provided, an alternative indicator of collusive pricing is price rigidity. Therefore, the coefficient of variation of ticket prices over the time can be used to measure the market-carrier pair level price rigidity, denoted as $CV_{hk,m}=\frac{\sigma_{hk,m}}{\mu_{hk,m}}$, in which $\sigma_{hk,m}$ and $\mu_{hk,m}$ are standard deviation and mean of the mean ticket prices of carrier $h$ and $k$ in market $m$ over the time. The average of pairwise coefficient of variation across observations is approximately $0.205$.

Therefore, the unit of observation in the regression is aggregated into either year-quarter-market-carrier pair or market-carrier pair. The second panel in table 3 exhibits descriptive statistics for the samples used in testing collusion via pairwise price difference, which contains 3,877,723 observations; the third panel displays that for samples used in testing collusion via pairwise coefficient of variation, which contains 217,886 observations. Providing the matching process and aggregation across the time, some variables, such as total number of passengers in the market, have different descriptive statistics across different panels.

Besides the introduction of a dummy variable to distinguish non-stop and stopover markets, one improvement of my paper is the introduction of controlling factors. Comparing with Ciliberto et al. (2019), number of passengers carried by all carriers in the market $m$ at time $t$, denoted as $TP_{mt}$, is introduced as a factor to control. The incentive of adding this controlling variable is because collusion in larger markets is more lucrative, while larger markets may have more competitors, making it difficult to achieve collusion. Therefore, adding market size represented by the number of passengers in the market as a control is necessary.

Another improvement of this paper is a more detailed measurement on market concentration beyond the Hirschmann-Herfindahl Index (HHI) used by Ciliberto et al. 2019 is also provided in this paper. This is because HHI only depicts the overall concentration level in the market despite the position of the carrier pair in the market, as at the same concentration level, carriers with dominant share are likely to behave differently from carriers at the fringe with little market share. Therefore, as this paper is using the time-market-carrier pair as the unit of observation in my regression, the expression of market concentration is bracketed down into two indicators, and both indicators range from $0$ to $1$. One is combined share, which is the sum of the market share of carrier $h$ and $k$ in market $t$ at time $t$, denoted as $CS_{hk,mt}=s_{h,mt}+s_{k, mt}$ if the market share of $h$ in market $m$ at time $t$ is denoted as $s_{h,mt}$. The other one is relative share, which is the market share of the carrier with less share in the carrier pair divided by the market share of the carrier with more share in the carrier pair. The relative share can be expressed as $RS_{hk,mt}=\frac{min(s_{h,mt}, s_{k,mt})}{max(s_{h,mt}, s_{k,mt}}$. 

\section{On Multimarket Contact and Collusion}

\subsection{Price Differences}
The first part of the paper is an extension from Ciliberto et. al. (2019), focusing on the relationship between collusive pricing and multimarket contact. The concept linking collusion with multimarket contact was first proposed by Bernheim and Whinston (1990). They have argued that when firms engage in multimarket contact, it combines the incentive constraints from all the markets they operate in. Essentially, the greater the overlap in the markets served by the two firms, the more significant the advantages of collusion and the penalties for breaking a collusive agreement become, thereby facilitating the maintenance of collusion.

The “baseline” regression from Ciliberto et. al. (2019) is:
\begin{center}
    $\Delta_{hk,mt}=\beta_{diff} \cdot (MMC^{EK}_hk,t)+\gamma_{diff}\mathbb{1}^{code-share}_{hk,t}+\alpha X_{mt}+FE_{hk,mt}+\epsilon_{hk,mt}$
\end{center}
In the regression, collusion is identified from the lack of price difference in differentiated goods. The intuition behind this setup is the services provided by each carrier are assumed to be differentiable goods. Such an assumption could be established with a multitude of reasons, such as service quality, loyalty and membership program, as well as alliance with other overseas airlines. Thus, if two carriers price too close, it implies a tacit collusion between them. Besides variables introduced above, $\mathbb{1}^{code-share}_{hk,t}$ is a dummy variable indicating whether carrier $h$ and $k$ have a code-sharing agreement at time $t$. $X_{hk,t}$ represents market-specific controlling variables, which are Hirschmann-Hirfindahl Index that measures market concentration of market $m$ at time $t$, and an dummy variable indicating the presence of low-cost carriers in market$m$ and time $t$. $FE_{hk,mt}$ indicates that fixed effects of city pairs, carrier pairs, year, and quarter are included.

The revised version used in this paper is:
\begin{center}
        $\Delta_{hk,mt}=\beta_{diff} \cdot (MMC^{EK}_hk,t)+\gamma_{diff}\mathbb{1}^{non-stop}_{hk,t}$\\
        $+\alpha_0 TP_{mt} + \alpha_1 CS_{hk, mt} + \alpha_2 RS_{hk, mt}+FE_{hk,mt}   +\epsilon_{hk,mt}$
\end{center}

The revised model has three main improvements. The first improvement is updated data to the second quarter of 2023. The second improvement is the distinction between non-stop and stopover flights, while previous literature categorizes markets merely by the origin and destination of the ticket itineraries. The third improvement is the introduction of more detailed controlling variables on market concentration.

In this revised regression, the hypothesis that multimarket contact lead to more collusive behavior can be testified if there is significant evidence that $\beta_{diff}<0$, and the regression supports the hypothesis that carriers are more collusive when pricing non-stop flights if there is significant evidence that $\gamma_{diff}<0$. On market and carrier-pair level controls, if there is $\alpha_{1}<0$, it can be inferred that two carrier are more likely to collude in a market which they have the majority of the share ; if there is $\alpha_{2}<0$, it can be interred that two carrier are more likely to collude in a market which they have a similar share. If there are both $\alpha_{1}<0$, $\alpha_{2}<0$, the hypothesis that carriers are most likely to collude in a duopoly market is supported.

Comparing to Ciliberto et. al. (2019), this paper has removed two independent varibles, which are the dummy indicating the presence of low-cost carriers in a market, and $\mathbb{1}^{code-share}_{hk,t}$ indicating the existence of code-sharing agreement between two carriers. The reason of removing the low-cost carriers dummy is the provision of basic economy option, which service is close to traditional low-cost carriers, by legacy carriers in domestic flights in the recent years, which has blurred the boundary between legacy and low-cost carriers. The removal of $\mathbb{1}^{code-share}_{hk,t}$ is due to the insignificance of its estimated coefficient in the regression result of Ciliberto et.al. (2019).

Provided the seasonality in demand indicated in Figure 1 and other factors of demand and cost varies across the time (such as fuel price) but constant over carrier pairs and city pairs, the year and quarter fixed effect is included in each regression. Furthermore, fixed effects coming from carrier pair and city pair\footnote{City pair rather than market is controlled as a fixed effect here is because a market is defined by city pair and non-stop dummy, and the latter is one of the independent variables. } have been controlled. This covers unobserved factors varies across either city pair or carrier pair, but remains constant over the time, such as the non-stop distance between a particular city pair. Moreover, the year-city pair fixed effect could be composited, which, besides original city pair fixed effect and year fixed effect, further absorbs all city pair-specific factors that varies changes in the long run on the yearly basis, such as the overall market size. Therefore, regressions using year-city pair fixed effect can exclude $TP_{m,t}$ as a market-level control.

\begin{table}[h!]
\centering
\small
\caption{Price Differences and Potential Facilitators of Collusion}
\vspace{12pt}
\begin{tabularx}{\textwidth}{l c c c c c c }
\hline
  & $(1)$ & $(2)$ & $(3)$ & $(4)$ & $(5)$ & $(6)$ \\
\hline
\multicolumn{7}{l}{\textit{Pairwise Covariates}}  \\
$MMC^{EK}$ & $-4.689^{***}$ & $-5.633^{***}$ &  &  &  &  \\
 & (0.04842) & (0.05077) &  &  &  &  \\
  &  &  &  &  &  &  \\
 $MMC^{CW}$ &  &  & $-22.226^{***}$ & $-33.280^{***}$ &  &  \\
 &  &  & (0.65984) & (0.68416) &  &  \\
  &  &  &  &  &  &  \\
 $MMC^{CW}_{Wgt}$ &  &  &  &  & $-60.632^{***}$ & $-29.591^{***}$ \\
 &  &  &  &  & (0.26267) & (0.35972) \\
  &  &  &  &  &  &  \\
 $\mathbb{1}^{non-stop}$ & $-1.698^{***}$ & $-7.912^{***}$ & $-1.480^{***}$ & $-7.682^{***}$ & $-2.025^{***}$ & $-7.740^{***}$ \\
 & (0.09963) & (0.08126) & (0.09972) & (0.08134) & (0.10077) & (0.09692) \\
  &  &  &  &  &  &  \\
 \multicolumn{7}{l}{\textit{Market and Carrier Pair-Level Controls}} \\
 $TP$ & $-0.001^{***}$ &  & $-0.002^{***}$ &  & $-0.002^{***}$ &  \\
 & (0.00002) &  & (0.00002) &  & (0.00003) &  \\
 $CombinedShare$ & $-2.374^{***}$ & $-8.647^{***}$ & $-2.858^{***}$ & $-9.256^{***}$ & $-1.702^{***}$ & $-10.069^{***}$ \\
 & (0.10089) & (0.10246) & (0.10092) & (0.10245) & (0.10010) & (0.08868) \\
 $RelativeShare$ & $-2.936^{***}$ & $-3.694^{***}$ & $-3.074^{***}$ & $-3.864^{***}$ & $-3.139^{***}$ & $-4.058^{***}$ \\
 & (0.08252) & (0.08313) & (0.08261) & (0.08325) & (0.08521) & (0.08627) \\
  &  &  &  &  &  &  \\
  \multicolumn{7}{l}{\textit{Fixed Effects}}  \\
  Year & Yes & No & Yes & No & Yes & No \\
  Quarter & Yes & Yes & Yes & Yes & Yes & Yes \\
  City Pair & Yes & No & Yes & No & Yes & No \\
  Carrier Pair & Yes & Yes & Yes & Yes & Yes & Yes \\
  Year-City Pair & No & Yes & No & Yes & No & Yes \\
 &  &  &  &  &  &  \\
   \multicolumn{7}{l}{\textit{Samples $\&$ Fit}}  \\
\# Carrier Pairs & 173 & 173 & 173 & 173 & 173 & 173 \\ 
\# City Pairs & 13,130 & 13,130 & 13,130 & 13,130 & 13,130 & 13,130 \\
\# Observations & 3,877,723 & 3,877,723 & 7,755,446 & 7,755,446 & 7,755,446 & 7,755,446 \\
Adjusted $R^2$ & 0.2312 & 0.2745 & 0.2296 & 0.2726 & 0.1741 & 0.2735 \\
 &  &  &  &  &  &  \\
\hline
\multicolumn{7}{l}{\textit{Note: Standard errors in parentheses: *** $p<0.01$, ** $p<0.05$, * $p<0.10$}}\\
\multicolumn{7}{l}{\textit{To keep measures of multimarket contact similar in scale, $MMC^{EK}_{hk,t}$ has been divided by 1000}}
\end{tabularx}
\end{table}

Alternatively, this paper estimates the regression using alternative definitions of multimarket contact, such as $MMC^{CW}_{h \rightarrow k,mt}$, which deflates $MMC^{CW}_{hk,mt}$ by the number of markets served by carrier $h$ at time $t$. $MMC^{CW}_{Wgt, h \rightarrow k,mt}$ is the passengers-weighted $MMC^{CW}_{h \rightarrow k}$, which is the total number of passengers carrier $h$ transported in markets both carriers $h$ and $k$ are present, divided by total number of passengers carrier $h$ transported.

Table 4 above presents the results from regressions between pairwise difference and multimarket contacts. Across the regressions, the adaptation of fixed effects and measures of multimarket contacts are varied. For $(2)$, $(4)$, and $(6)$, the composited year-city pair fixed effect is adapted, which captures city pair-specific factors that varies changes in the long run on the yearly basis, which includes overall market size. Therefore, $TP_{m,t}$ are not in the independent variables of these regressions.

Using $MMC^{EK}_{hk,t}$ as the multimarket contact measure, column 1 of table 4 provides a baseline regression. It estimates that $\beta_{diff}=-4.689$ and is statistically significant, which implies that $1000$ more multimarket contact is associated with a $\$4.689$ decrease in the pairwise price difference in mean ticket prices of two carriers. Also, it has a significant estimation that $\gamma_{diff}=-1.698$, associating non-stop markets with a smaller pairwise price difference, implying that carriers are more collusive in non-stop markets.

In column 2, adapting a different approach of fixed effect control, a similar statistically significant estimation $\beta_{diff}=-5.644$ is attained. Notably, the magnitude of the coefficient on $\gamma_{diff}$ has exaggerated. Indicating that when controlling the overall market size as part of the year-city pair fixed effect, the effect of non-stop dummy on the pairwise pair difference could be more significantly identified, fortified the observation that carriers are more collusive in non-stop markets.

Column 3 and 4 of Table 4 presents the results from regressions using $MMC^{CW}_{h \rightarrow k,t}$, and column 5 and 6 uses $MMC^{CW}_{Wgt, h \rightarrow k,mt}$ For all four regressions, this paper find that with estimated $\hat{\beta_{diff}}<0$, implying that they are qualitatively coherent with the previous regression. The numerical difference in the estimated coefficient could be partially explained by the scale difference in different measures of multimarket contact (after divided by 1,000, $MMC^{EK}_{hk,t}$ ranges from 0.001 to 6.097, while other two measures range from 0.002 to 1). Similarly, all four regressions have derived an estimated coefficient on  $\gamma_{diff}$ that is negative and statistically significant, which further implies that carriers are more collusive in non-stop market.

For the estimated coefficients on market and carrier-pair level controls, across all specifications, significant, negative estimated $\alpha_0$ could be observed on regression $(1)$, $(3)$, and $(5)$. This provides an association between larger market size and smaller pairwise price difference, as it estimates that every $1000$ more passengers transported in a market per quarter is connected to a $\$1$ decrease in the pairwise price difference in mean ticket prices of two carriers. Moreover, across all specifications, there are significant statistical evidence that ${\alpha_1}<0$ and ${\alpha_2}<0$. This associates year-quarter-market-carrier pairs that have higher combined share and relative share with lower pairwise difference, which is an indicator of collusion, and thereby supports the hypothesis that carriers are most likely to collude in a duopoly market. Notably, estimated $\alpha_2$ are significantly larger in magnitude in regression $(2)$, $(4)$, and $(6)$, indicates that introducing composite year-city pair fixed effect to control, such correlation is amplified and easier to observe.

To conclude, the results in table 4 reveals a negative relationship between multimarket contact and pairwise price difference, supporting the hypothesis that multimarket contact facilitates collusive pricing between airlines. Furthermore, the results support the hypothesis that airlines are more collusive when pricing non-stop flights. With a negative relationship between combined share, relative share, and the pairwise price difference, the results also supports the hypothesis that carriers are most likely to collude in a duopoly market. 

\subsection{Price Rigidity}

There are alternative choices of dependent variables to identify collusion. For example, using the conclusion from Athey et.al.(2004), beyond absolute and relative price level, price rigidity could be also utilized as a means to identify collusion. The theoretical underpinnings behind that is collusive firms have incentive to their prices steady despite changes in costs or demand to maintain oligopolistic order. Athey et.al.(2004) further concludes that steadfastness in pricing helps balance the efficiency gains from redistributing market shares following privately observed changes in costs against the information costs colluding firms encounter in detecting price reductions among competitors.

To test the relationship between multimarket contact and collusion using price rigidity as a signal, this paper adapts the following regression:
\begin{center}
        $CV_{hk,m}=\beta_{se} \cdot (MMC^{EK}_{hk,2023Q2})+\gamma_{se}\mathbb{1}^{non-stop}_{hk,t}$\\
        $ + \alpha_1 CS_{hk, m} + \alpha_2 RS_{hk, m}+FE_{hk,m}   +\epsilon_{hk,m}$    
\end{center}

Notably, $CV_{hk,m}$ is the coefficient of variation of mean ticket prices of the two carriers $h$ and $k$ in market $m$ over the time. Therefore, in this regression, $CS_{hk,m}$ and $RS_{hk,m}$ are also calculated by the sum of the number of passengers transported by carrier $h$ and $k$ over the time. As the dimension of time no longer exist in this model above, the overall market size has become a city pair-varied but carrier pair-fixed factor, which can be absorbed when controlling city pair-fixed effects. Therefore, $TP_{m,t}$in the previous part, or the total number of passengers at all times, is not included in this part of the paper. For the multimarket contact, the newest statistic from the second quarter of 2023 is used in this regression.

\begin{table}[h!]
\centering
\small
\caption{Price Rigidity and Potential Facilitators of Collusion}
\vspace{12pt}
\begin{tabularx}{\textwidth}{ l Y Y Y }
\hline
  & $(1)$ & $(2)$ & $(3)$ \\
\hline
\multicolumn{4}{l}{\textit{Pairwise Covariates}}  \\
$MMC^{EK}$ & $-0.002^{***}$ &  &  \\
  & (0.00016) &  & \\
  &  &  & \\
 $MMC^{CW}$ &  & $-0.042^{***}$ &  \\
  &  & (0.00203) &  \\
  &  &  &  \\
 $MMC^{CW}_{Wgt}$ &  &  & $-0.002^{***}$ \\
 &  &  & (0.00132) \\
  &  &  &  \\
 $\mathbb{1}^{non-stop}$ & $0.004^{***}$  & $0.004^{***}$ & $0.004^{***}$ \\
 & (0.00079) & (0.00079) & (0.00079) \\
  &  &  &  \\
 \multicolumn{4}{l}{\textit{Market and Carrier Pair-Level Controls}} \\
 $CombinedShare$ & $0.051$ & $-0.040$ & $0.061$ \\
 & (0.04161) & (0.04160) & (0.04164) \\
 $RelativeShare$ & $0.0014$ & $-0.002^{}$ & $0.001$ \\
 & (0.00101) & (0.08578) & (0.00102) \\
  &  &  & \\
  \multicolumn{4}{l}{\textit{Fixed Effects}}  \\
  City Pair & Yes & Yes & Yes \\
  Carrier Pair & No & No & No \\
 &  &  & \\
   \multicolumn{4}{l}{\textit{Samples $\&$ Fit}}  \\
\# Carrier Pairs & 173 & 173 & 173 \\ 
\# City Pairs & 13,130 & 13,130 & 13,130 \\
\# Observations & 217,886 & 435,772 & 435,772 \\
Adjusted $R^2$ & 0.1212 & 0.1222 & 0.1204 \\
 &  &  &  \\
\hline
\multicolumn{4}{l}{\textit{Note: Standard errors in parentheses: *** $p<0.01$, ** $p<0.05$, * $p<0.10$}}
\end{tabularx}
\end{table}

In this regression, the unit of observation is at market-carrier pair level, and the database has collapsed into a cross-sectional data. Besides removing all time-varied fixed effects, such a data structure has also made a two-way fixed effect(TWFE) regression not possible. Without any other dimension, there's no within-entity variation to exploit, making it impossible to apply two sets of entity-specific fixed effects. Essentially, with only one or two observations per entity, all variability that might be attributed to entity-specific effects from two dimensions would be perfectly collinear with these fixed effects, leaving no variation to estimate the impact of other independent variables. Therefore, when regressing price rigidity on multimarket contacts, only city pair fixed effect is adapted. Moreover, when 

Similarly, the hypothesis that multimarket contact lead to more collusive behavior can be testified if there is significant evidence that $\beta_{se}<0$. In table 5, three different regressions with the same set of control variables and fixed effect, but with different measurements of multimarket contacts, are presented. Column 1 of the table 5 reveals the estimation of $\beta_{se}$ when $MMC^{EK}_{hk, 2023Q2}$ is the measure of multimarket contact. The regression estimates $\beta_{se}=-0.002$, and is statistically significant. Column 2 and 3 regress $CV_hk,m$ on $MMC^{CW}_{hk, 2023Q2}$ and $MMC^{CW}_{Wgt,hk, 2023Q2}$ respectively, which both attain a negative, significant estimation of $\beta_{se}$. Therefore, the estimated $\beta_{se}$ in all three regressions displayed in table 5 support the hypothesis. 

Notably, the estimated $\alpha_1$ and $\alpha_2$ are insignificant across all three regressions in table 5, and are positive in column 1 and 3. This implies that there is lack of evidence on the association between the pairwise relative share as well as the combined share of two carrier and the coefficient of variation of their mean ticket prices over the time. This may because the carrier pair fixed effect is not included in the regression due to the data structure. Moreover, The estimation on $\gamma_{se}$ is also positive and statistically significant, but very small in magnitude, indicating that the coefficient of variation of the mean ticket prices over the time are 0.004 higher in non-stop markets comparing with stopover markets. The estimated coefficient above indicates that prices are less rigid in non-stop flights. However, besides collusive pricing, price rigidity also associates with seasonality of demand, which is uncontrolled because of absence of the time dimension in the regressions above; lack of price rigidity can also be explained as a result of price discrimination, which are common in airlines industry. Therefore, the estimated coefficient on $\gamma_{se}$ provides another facet of the pricing dynamic of airlines, but does not conflict with the conclusion made in the previous part regarding market concentration, non-stop flights, and level of collusive pricing.

\section{On Merger and Collusive Pricing}

The second part of the paper examines the impact of airline mergers on collusion. Considering the time magnitude of a merger, in this part, This paper uses event study method in alignment with Goolsbee and Syverson (2008) measure the impact of a particular merger in different stages, and applies difference-in-difference element to examine the asymmetric effect of the merger on the collusion between different sets of carrier pairs.

\begin{center}
        $\Delta_{hk,mt}=\beta_{diff} \cdot (MMC^{EK}_hk,t)+\gamma_{diff}\mathbb{1}^{non-stop}_{hk,t}$\\
        $+\theta_{11}\mathbb{1}_{annc}+\theta_{12}\mathbb{1}_{annc}\cdot\mathbb{1}_{either}+\theta_{13}\mathbb{1}_{annc}\cdot\mathbb{1}_{both}$\\
        $+\theta_{21}\mathbb{1}_{appro}+\theta_{22}\mathbb{1}_{appro}\cdot\mathbb{1}_{either}+\theta_{23}\mathbb{1}_{appro}\cdot\mathbb{1}_{both}+\theta_{31}\mathbb{1}_{cert}+\theta_{32}\mathbb{1}_{cert}\cdot\mathbb{1}_{either}$\\
        
        $+\alpha_1 TP_{mt} + \alpha_2 CS_{hk, mt} + \alpha_3 RS_{hk, mt}+FE_{hk,mt}   +\epsilon_{hk,mt}$
\end{center}

Besides multimarket contacts as well as market and carrier-pair controls specified in the previous parts, two arrays of treatment and time dummies are adapted, which are time dummies $\mathbb{1}_{annc}$, $\mathbb{1}_{appro}$, and $\mathbb{1}_{cert}$, as well as treatment dummies $\mathbb{1}_{either}$ and $\mathbb{1}_{both}$.

The design of time dummies results from the characteristics of airline merger. Usually, the merger initiative is disclosed when the shareholders and executives of merged carriers have reached an agreement.Then, the merger process is subject to regulatory approval, antitrust review, as well as bargaining with the labor union, which takes months and even years. After regulatory approval, the integration process of operation, reservation, and membership system of the merged carrier will begin. Upon the completion of this process, a single operating certificate for the merged carrier, which officially summarizes the merger, is granted. Besides the completion of the merger, the agreement of shareholders, as well as the completion of antitrust review, could incentivize carriers to alternate their pricing strategies. Therefore, to evaluate the impact of airline merger over the time, three time dummies corresponding to three event windows are created: $\mathbb{1}_{annc}$ implies after the announcement of the merger and before the regulatory approval; $\mathbb{1}_{appro}$ implies after the regulatory approval and before the single operating certificate is granted; and $\mathbb{1}_{cert}$ implies after the single operating certificate is granted.

The treatment dummies are created to explore the asymmetric effect of the merger on the pairwise price difference between two merged carrier, a merged carrier and a unmerged carrier such as United and American in the United-Continental merger, and two unmerged carriers such as American and Jetblue in the United-Continental merger. Therefore, two treatment dummies are created: $\mathbb{1}_{either}$ indicates either (but not both) carriers in the year-quarter-market-carrier pair level observation is one of the merged carriers; $\mathbb{1}_{both}$ indicates two carriers in the year-quarter-market-carrier pair level observation is the two merged carriers. Note that as the single operating certificate represents the completion of the merger process, $\mathbb{1}_{cert} \cdot \mathbb{1}_{both}=0$ for all observations as the brand, code, and callsign of one of the merged carriers perishes when the merger is completed, and thus will not be found in later observations.

Specifically, This paper focuses on two merger cases. The first one is the merger between United Airlines and Continental Airlines. It was announced on 3 May 2010 and smoothly passed through antitrust overview on 27 August in the same year due to minimal overlap in their routes. Merged operating certificate was granted on 30 November 2011. The second case is the merger between American Airlines (AA) and U.S. Airways (US). The merger was officially announced on 14 February 2013, and was objected to by the U.S. Department of Justice and six states, which filed a lawsuit seeking to block the merger. In the settlement attained on 12 November in the same year, the merged airline was required to give up landing slots or gates in 7 major airports. FAA granted operating certificate to the merged carrier on 8 April 2015. Given the apparently different attitude of regulatory agencies towards two merger cases, evaluating the effect on collusive pricing from each merger will be a helpful review and evaluation on the antitrust practice in the airline industry. 

With the regression design above, for the gross effect of merger on the collusive level, statistically significant evidence showing $\theta_{11}<0$, $\theta_{21}<0$, and $\theta_{31}<0$ supports hypothesis that in the respective stages, merger leads to more collusion among unmerged carriers. Likewise, statistically significant evidence indicating $\theta_{11}+\theta_{12}<0$, $\theta_{21}+\theta_{22}<0$, $\theta_{31}+\theta_{32}<0$ supports hypothesis that in the respective stages, merger leads to more collusion between the merged carrier and unmerged carrier. Moreover, a estimation that $\theta_{11}+\theta_{13}<0$ implies after the announcement of the merger initiative, two carriers involved in the merger procedure become more collusive in pricing; $\theta_{21}+\theta_{23}<0$ indicates that after regulatory approval, two carriers involved in the merger procedure become more collusive.

Furthermore, the asymmetry of the impact from merger on the collusion between different types of carrier pairs can be detected. A statistically significant evidence showing $\theta_{21}<0$, $\theta_{22}<0$, and $\theta_{23}<0$ supports the hypothesis that in the respective stages, collusion are relatively more likely to take place between a merged carrier and a unmerged carrier comparing with two unmerged carriers. Statistically significant evidence showing $\theta_{31}<0$, $\theta_{32}<0$ supports the hypothesis that in the respective stages before the completion of the merger process, two merged carriers are more collusive than two unmerged carriers. 
\begin{table}[!htbp]
\centering
\small
\caption{Pairwise Price Differences and Mergers}
\vspace{12pt}
\begin{tabularx}{\textwidth}{l c c c c c c }
\hline
  & $(1)$ & $(2)$ & $(3)$ & $(4)$ & $(5)$ & $(6)$ \\
\hline
\multicolumn{7}{l}{\textit{Multimarket Contact}}  \\
$MMC^{EK}$ & $-4.526^{***}$ &  &  & $-4.857^{***}$ &  &  \\
 & (0.02149) &  &  & (0.02359) &  &  \\
  &  &  &  &  &  &  \\
 $MMC^{CW}$ &  & $-49.234^{***}$ &  &  & $-48.286^{***}$ &  \\
 &  & (0.65984) &  &  & (0.27152) &  \\
  &  &  &  &  &  &  \\
 $MMC^{CW}_{Wgt}$ &  &  & $-18.850^{***}$ &  &  & $-12.868^{***}$ \\
 &  &  & (0.19613) &  &  & (0.19316) \\
  &  &  &  &  &  &  \\
 \multicolumn{7}{l}{\textit{Post-Annoucement Pre-Regulatory Approval}} \\
 $\mathbb{1}_{annc}$ & $0.803^{***}$ & $0.908^{***}$ & $0.941^{***}$ & $8.702^{***}$ & $6.598^{***}$ & $6.451^{***}$ \\
 & (0.29943) & (0.29967) & (0.30067) & (0.20790) & (0.20790) & (0.20863) \\
 $\mathbb{1}_{annc} \cdot \mathbb{1}_{either}$ & $-2.041^{***}$ & $-0.641^{}$ & $-0.982^{**}$ & $0.208^{}$ & $-0.993^{***}$ & $-0.335^{}$ \\
 & (0.44037) & (0.44096) & (0.44250) & (0.27602) & (0.27662) & (0.27773) \\
 $\mathbb{1}_{annc} \cdot \mathbb{1}_{both}$ & $-7.950^{***}$ & $-0.472^{***}$ & $-3.452^{***}$ & $-1.687^{***}$ & $-0.818^{***}$ & $-0.124^{***}$ \\
 & (1.19417) & (1.19563) & (1.20008) & (0.64733) & (0.64821) & (0.65066) \\
  &  &  &  &  &  &  \\
 \multicolumn{7}{l}{\textit{Post Regulatory Approval Pre-Operating Certificate Merger}} \\
 $\mathbb{1}_{appro}$ & $2.434^{***}$ & $0.588^{***}$ & $0.651^{***}$ & $16.428^{***}$ & $13.989^{***}$ & $13.601^{***}$ \\
 & (0.13584) & (0.13569) & (0.13615) & (0.15151) & (0.15105) & (0.15157) \\
 $\mathbb{1}_{appro}\cdot \mathbb{1}_{either}$ & $-1.990^{***}$ & $-0.487^{**}$ & $-0.594^{***}$ & $-6.288^{***}$ & $-5.378^{***}$ & $-6.904^{***}$ \\
 & (0.20175) & (0.20207) & (0.20292) & (0.19686) & (0.19747) & (0.19819) \\
 $\mathbb{1}_{appro} \cdot \mathbb{1}_{both}$ & $-11.124^{***}$ & $-6.352^{***}$ & $-3.771^{***}$ & $-17.623^{***}$ & $-15.839^{***}$ & $-13.855^{***}$ \\
 & (0.52222) & (0.52332) & (0.52932) & (0.40755) & (0.40829) & (0.41445) \\
  &  &  &  &  &  &  \\
 \multicolumn{7}{l}{\textit{Post-Operating Certificate Merger}} \\
 $\mathbb{1}_{cert}$ & $10.586^{***}$ & $9.942^{***}$ & $10.478^{***}$ & $14.960^{***}$ & $12.556^{***}$ & $21.678^{***}$ \\
 & (0.06130) & (0.05638) & (0.05643) & (0.07001) & (0.06819) & (0.06830) \\
 $\mathbb{1}_{cert} \cdot \mathbb{1}_{either}$ & $-7.797^{***}$ & $-6.68^{***}$ & $-7.724^{***}$ & $-9.417^{***}$ & $-12.662^{***}$ & $-16.12151^{***}$ \\
 & (0.07550) & (0.07597) & (0.07594) & (0.09617) & (0.09219) & (0.09033) \\
  &  &  &  &  &  &  \\
  \multicolumn{7}{l}{\textit{Fixed Effects}}  \\
  Quarter & Yes & Yes & Yes & Yes & Yes & Yes \\
  City Pair & Yes & Yes & Yes & Yes & Yes & Yes \\
 &  &  &  &  &  &  \\
   \multicolumn{7}{l}{\textit{Samples $\&$ Fit}}  \\
\# Carrier Pairs & 173 & 173 & 173 & 173 & 173 & 173 \\ 
\# City Pairs & 13,130 & 13,130 & 13,130 & 13,130 & 13,130 & 13,130 \\
\# Observations & 3,877,723 & 7,755,446 & 7,755,446 & 3,877,723 & 7,755,446 & 7,755,446 \\
Adjusted $R^2$ & 0.1440 & 0.1420 & 0.1363 & 0.1587 & 0.1741 & 0.1018 \\
 &  &  &  &  &  &  \\
\hline
\multicolumn{7}{l}{\textit{Note: Standard errors in parentheses: *** $p<0.01$, ** $p<0.05$, * $p<0.10$}} \\
\multicolumn{7}{l}{\textit{Market and carrier pair-level controls included in the regressions on price differences are also}} \\
\multicolumn{7}{l}{\textit{included here. There coefficients are in the appendix.}} \\
\end{tabularx}
\end{table}

Table 6 displays the result from the regression estimating the impact of mergers on pairwise price differences, which is a signal of collusive pricing. As the treatment dummies are carrier pair-varied and fixed in all other dimensions and the time dummies are collinear with unobserved year fixed effects, we will be unable to control the year and carrier pair fixed effects in this group of regressions. 

Column 1 to 3 present the results from regressions using time and treatment dummies constructed from the United Airlines-Continental Airlines merger. Regression displayed in column 1 adapts $MMC^{EK}_{hk,mt}$ as the measure of multimarket contacts. It has a estimated $\beta_{diff}=-4.526$ and is statistically significant, which, as well as estimated coefficients on other controlling variables introduced, aligns with the estimation displayed in the previous part of this paper. Interestingly, the regression estimates $\theta_{11}>0$, $\theta_{21}>0$, as well as $\theta_{31}>0$, and all estimated coefficients are statistically significant. This supports the contrary of the hypothesis mentioned before, implying that merger may associate with less collusion among unmerged carriers. 

The estimated effect of merger on the pairwise price difference between a merged carrier and a unmerged carrier by the model is $\theta_{11}+\theta_{12}$, $\theta_{21}+\theta_{22}$, and $\theta_{31}+\theta_{32}$ for respective stages. In the regression displayed in column 1, $95\%$ CI for $\theta_{11}+\theta_{12}$ is $[-2.282, -0.194]$; $95\%$ CI for $\theta_{21}+\theta_{22}$ is $[-0.033, 0.921]$, and $95\%$ CI for $\theta_{31}+\theta_{32}$ is $[-2.598, 2.980]$. Therefore, there is no statistically significant evidence that the pairwise price difference between a merged carrier and a unmerged carrier decreases when the merger process is completed. Therefore, the regression outcome suggests there is no sufficient evidence that after the United-Continental merger takes place, airlines become more collusive than before.

However, the regression estimates that $\theta_{12}<0$, $\theta_{22}<0$, as well as $\theta_{32}<0$, which supports the hypothesis that throughout and after the merger process, the price difference in a pairwise observation between a merged carrier and a unmerged carrier is less than that in a pairwise observation between two unmerged carriers. This implies a higher collusive level between a merged and an unmerged carrier comparing with two unmerged carriers. In the context of United-Continental merger, if the pairwise price difference for American-United and American Delta in a particular market are identical before the merger, the estimated coefficient $\theta_{32}<0$ predicts that once the merger was completed, the price difference of American-United pair will be smaller than that of American-Delta pair. Such an observation may result from the increasing market power of the merged airlines. 

Ciliberto et. al. (2014) asserts that the impact of airline merger on collusive pricing could be adequately explained by the change in multimarket contact. However, effects from increasing multimarket contact are controlled by the multimarket contact statistics. The effects coming from an alternation in market power due to the change of market share in a particular market are also controlled by the market-carrier pair level combined and relative share statistics. Therefore, the estimated coefficient above implies there exists an alternative mechanism that an airline merger could affect the collusive pricing of carriers. One potential explanation for such a dynamic is an increase in network externality resulted from the merger. After the merger, the service network of the merged airline expands, resulting in a higher network externality that increase the market power of the merged carrier, and hence facilitates collusion \footnote{An expansion in network due to merger does not necessarily increases the multimarket contact, as some service are (or become) uncontested after the merger}. 

Column 2 and column 3 present the regression results using same dummies but with alternative measurements of multimarket contacts. Results from both regressions are coherent with the result from the regression presented above. The estimated coefficient on $\theta_{11}$, $\theta_{21}$, and $\theta_{31}$ does not support the hypothesis that merger facilitates more collusion among unmerged carriers, while the estimated coefficient on $\theta_{32}$ implies that when the merger is completed, if other controlling factors are identical, carrier pairs between merged and unmerged carriers are likely to be more collusive comparing with carrier pairs with unmerged carriers.

Notably, all three models provide statistically significant estimations that $\theta_{13}<0$ and $\theta_{23}<0$, as well as $\theta_{11}+\theta_{13}<0$ and $\theta_{21}+\theta_{23}<0$. This implies that once the United-Continental merger initiative was announced, the pairwise price difference between them has shrunk, which is a signal of collusion. This could be a piece of evidence that prior to regulatory approval, United and Continental in the merger process has begun to price collusively.

Column 4 to 6 present the result from regressions using time and treatment dummies constructed from the American Airlines-U.S. Airways merger, in which a pattern similar to the United-Continental merger is exhibited. First, there is no sufficient statistical evidence supporting the hypothesis that merger facilitates more collusion among unmerged carriers or between merged and unmerged carriers. However, there is statistical evidence supporting that during and after the merger process, carrier pairs between merged and unmerged carriers are more likely to exhibit collusive pricing comparing with carrier pairs with unmerged carriers. 

Different from the United-Continental merger, there is no sufficient statistical evidence on either $\theta_{11}+\theta_{13}<0$ or $\theta_{21}+\theta_{23}<0$ when examine the model on American-U.S. Airways merger. This suggests that prior to the formal completion of American -U.S. Airways merger, the pairwise price difference between them, which is a indicator of collusion between them, does not significantly decrease. Such a difference may result from the uncertainty aroused from the initial objection from the regulatory authorities, in which U.S. Department of justice, along with attorneys general from six states and Washington D.C., filed a lawsuit to block the merger in August 2013. Such a move could make both merged carrier more prudent in tacit collusion before the merger process was completed. 

\section{Conclusion}

This comprehensive study on the dynamics of multimarket contact, airline mergers, and their implications for collusive pricing confirms that increased multimarket contact is associated with reduced price differences among airlines, indicating a collusive equilibrium facilitated by frequent interactions across multiple markets. This aligns with the theories proposed by Bernheim and Whinston (1990), which suggest that extensive multimarket contact can enhance the sustainability of collusion by increasing the punitive costs of competitive deviations. The empirical findings of this paper support this theory, showing a consistent negative relationship between multimarket contact and price dispersion. To ensure the robustness of these results, we also employed several alternative measurements of multimarket contacts and an auxiliary empirical strategy using price rigidity to identify collusion. The estimated coefficients from these alternative models corroborate the finding that multimarket contact facilitates collusive pricing between carriers.

Moreover, our empirical strategy distinguishes between non-stop and stopover markets and introduces more detailed control variables on market concentration. Using pairwise price differences as a signal of collusive pricing, we reveal that airlines price tickets more collusively for non-stop flights, where there are fewer competitors and less price-sensitive consumers. Additionally, we find a relationship between market share and collusive pricing: the closer two carriers are to achieving a duopoly in a pairwise observation, the more likely they are to price collusively.

Our findings present a nuanced view of the relationship between airline mergers and collusive pricing. Despite changes in multimarket contact, airline mergers still affect collusive pricing. Although it is insufficient to directly conclude that carriers price more collusively after a merger than before, the empirical results show that during and after the merger process, carrier pairs between merged and unmerged carriers are more likely to collude compared to pairs of unmerged carriers. This phenomenon may be explained by an increase in network externalities resulting from the merger. Specifically, for the United-Continental case, it is revealed that as soon as the merger process begins, the pairwise price difference between the two merged carriers decreases, implying a likelihood of collusive pricing.

This thesis contributes to the understanding of how multimarket contact and airline mergers influence collusive behavior in the airline industry. It not only reaffirms established theories but also extends the discussion to include non-stop versus stopover flights, market concentration, and airline mergers. This study also provides a useful review of antitrust practices, especially when comparing the United-Continental merger with the American-U.S. Airways merger.

However, several aspects of this paper require further elaboration. Firstly, the intuition behind the observation that two carriers are more likely to collude when they are closer to achieving a duopoly suggests that carriers may engage in Cournot monopolistic competition when setting prices and quantities, but further theoretical deduction is needed. Secondly, the relationship between network externalities and collusive pricing requires additional exploration. Even when controlling for multimarket contacts and market concentration, airline mergers still affect collusive pricing, presumably through increased network externalities of the merged carriers. This interpretation warrants further investigation in future research.

\section{Appendix: Collinearity Tests of Variables}

\begin{figure}[ht]
    \centering
    \begin{subfigure}[b]{0.32\textwidth}
    \includegraphics[width=\textwidth]{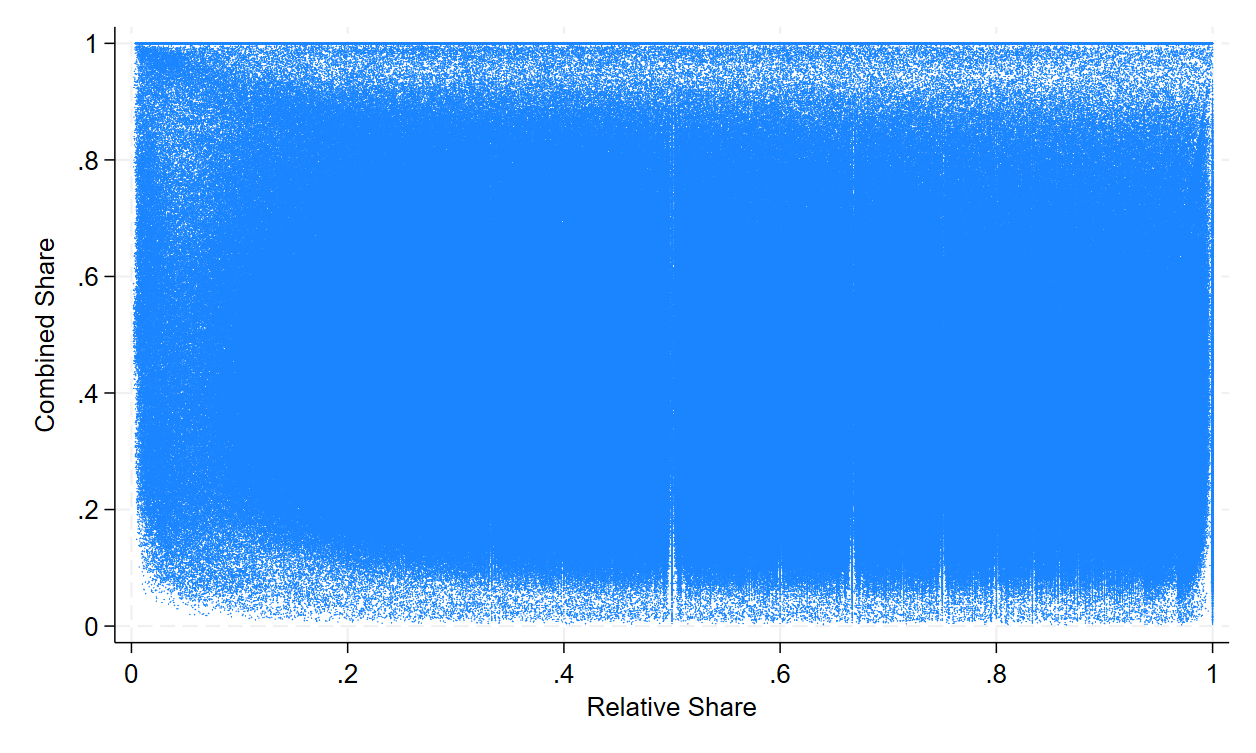}
    \caption{$CS$ and $RS$(-0.04)}
    \end{subfigure}
    \begin{subfigure}[b]{0.32\textwidth}
    \includegraphics[width=\textwidth]{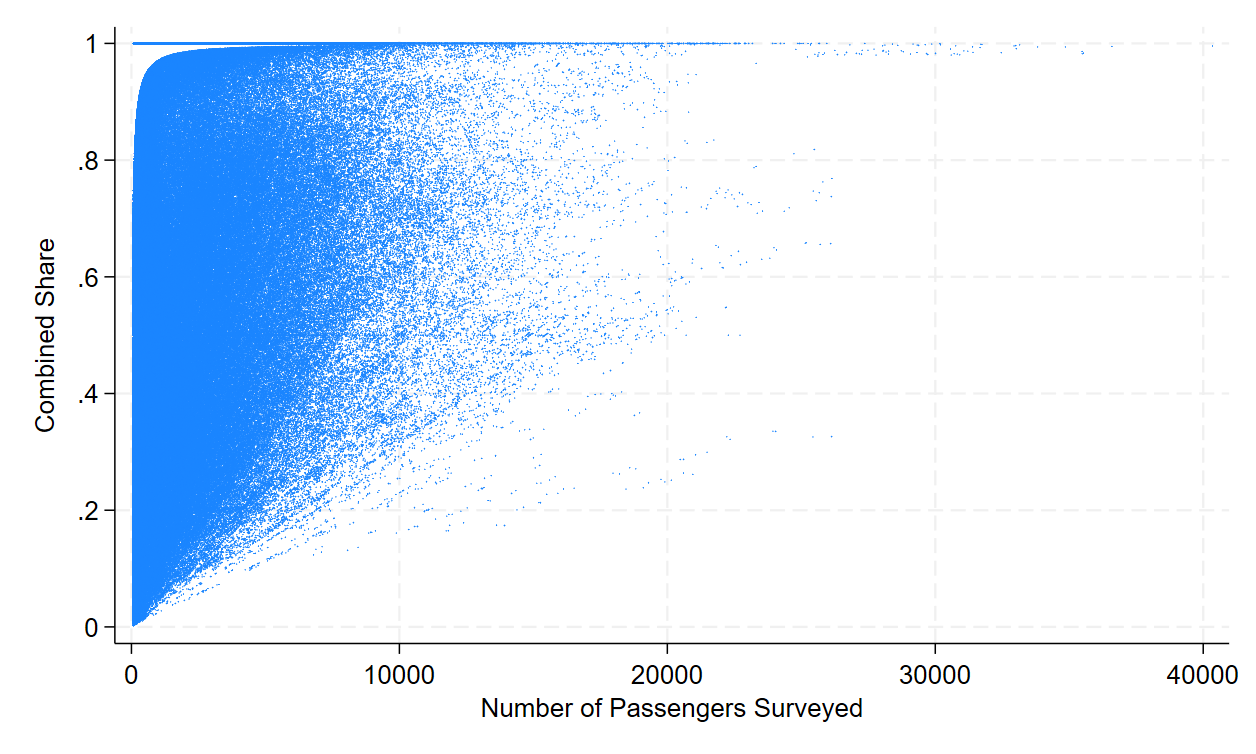}
    \caption{$CS$ and $TP$(0.17)}
    \end{subfigure}
    \begin{subfigure}[b]{0.32\textwidth}
    \includegraphics[width=\textwidth]{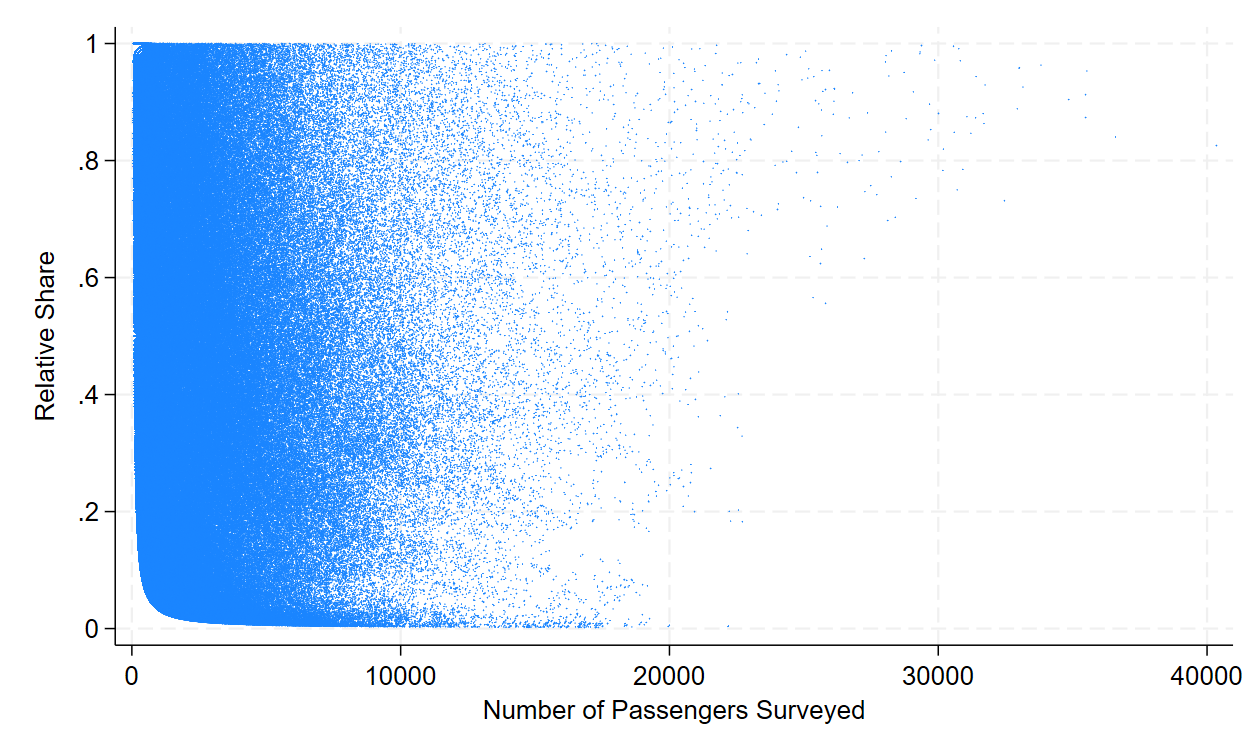}
    \caption{$RS$ and $TP$(-0.15)}
    \end{subfigure}
    \centering
    \begin{subfigure}[b]{0.32\textwidth}
    \includegraphics[width=\textwidth]{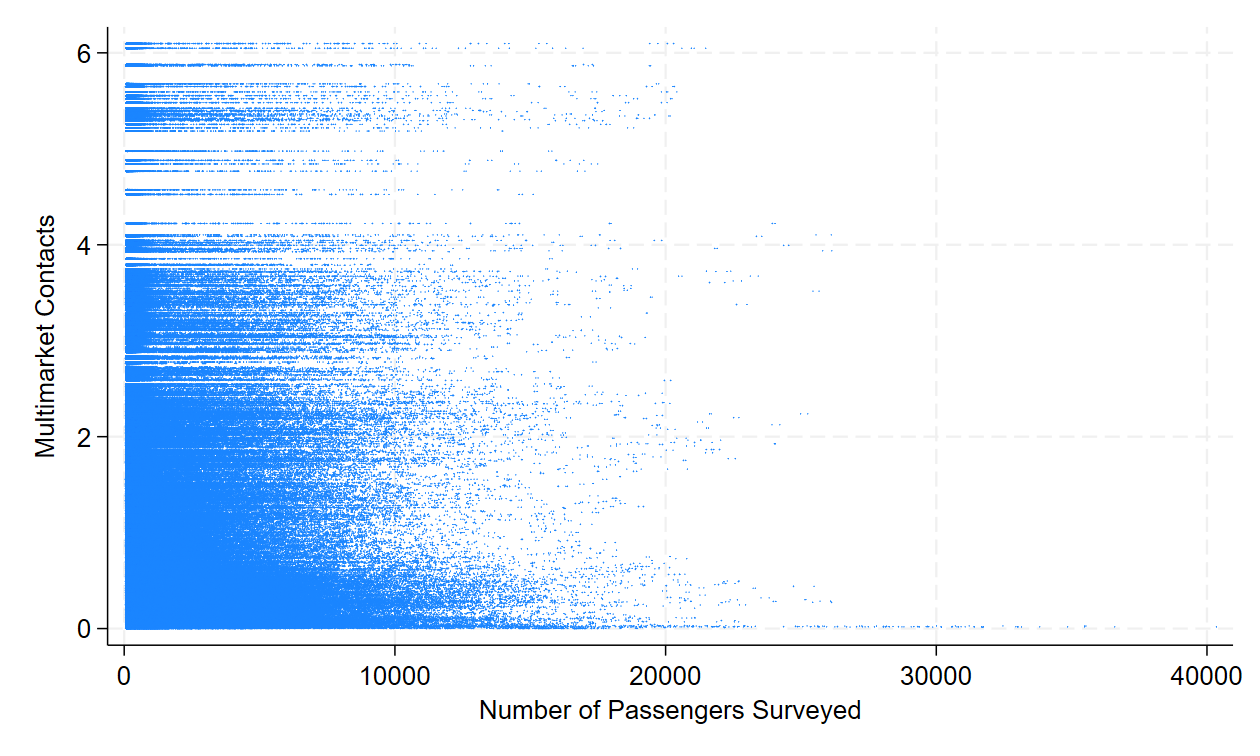}
    \caption{$MMC^{EK}$ and $TP$(-0.09)}
    \end{subfigure}
    \begin{subfigure}[b]{0.32\textwidth}
    \includegraphics[width=\textwidth]{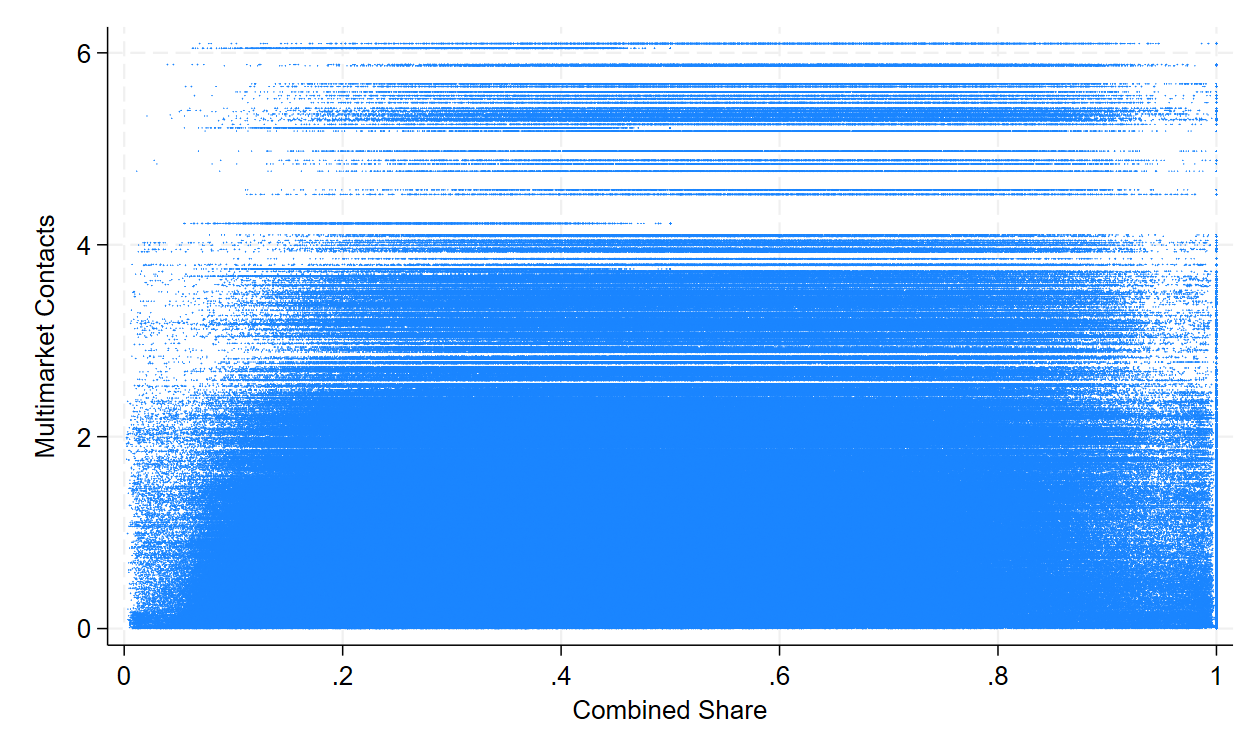}
    \caption{$MMC^{EK}$ and $CS$(0.26)}
    \end{subfigure}
    \begin{subfigure}[b]{0.32\textwidth}
    \includegraphics[width=\textwidth]{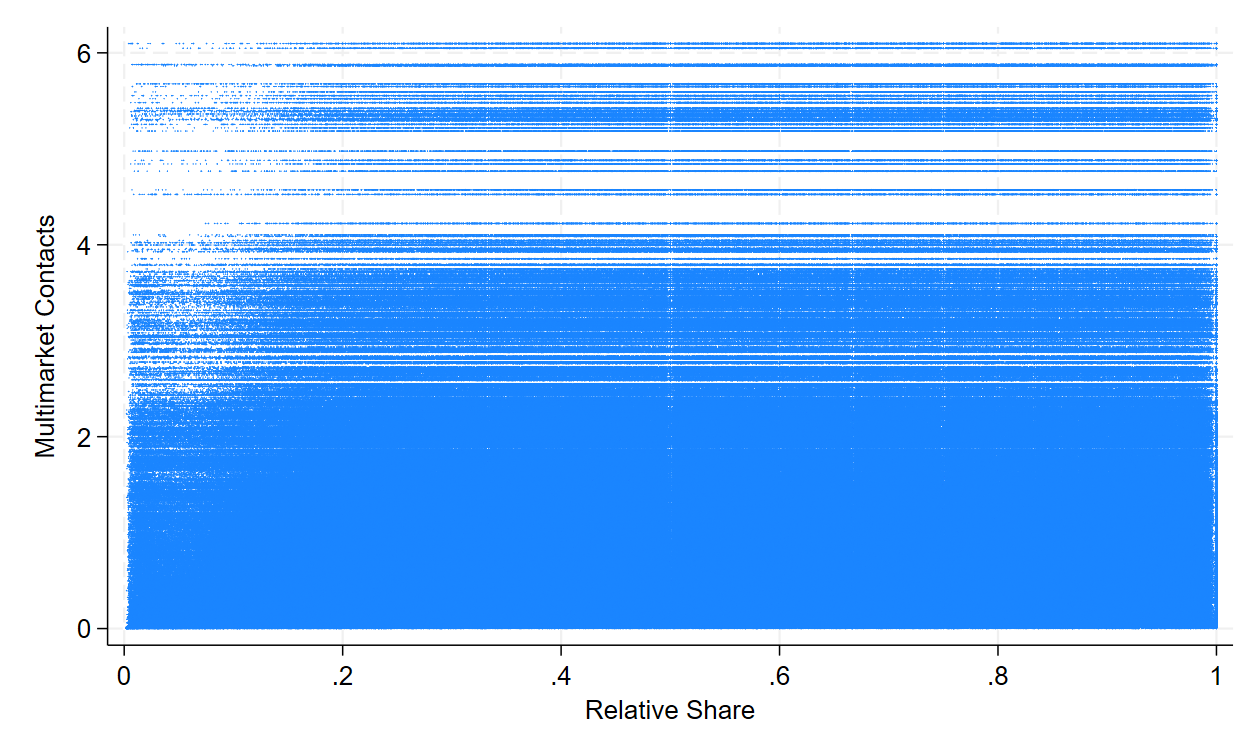}
    \caption{$MMC^{EK}$ and $RS$(0.11)}
    \end{subfigure}
    \centering
    \begin{subfigure}[b]{0.32\textwidth}
    \includegraphics[width=\textwidth]{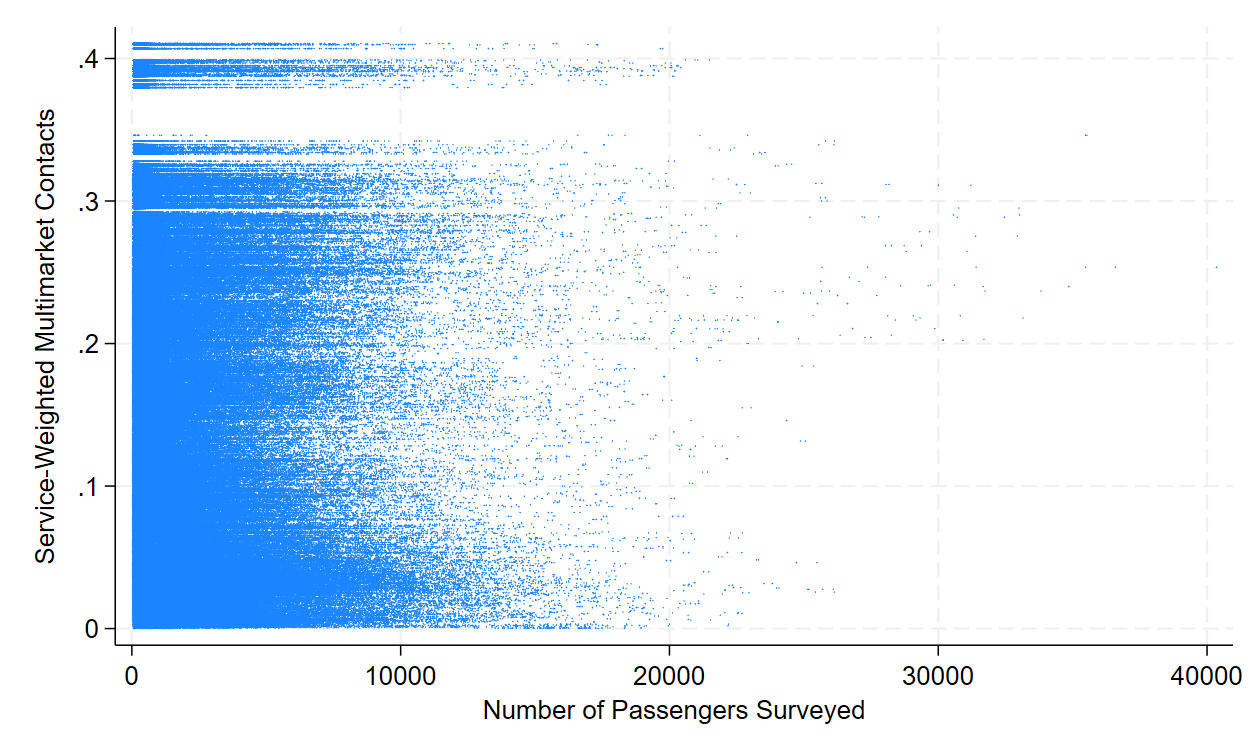}
    \caption{$MMC^{CW}$ and $TP$(-0.13)}
    \end{subfigure}
    \begin{subfigure}[b]{0.32\textwidth}
    \includegraphics[width=\textwidth]{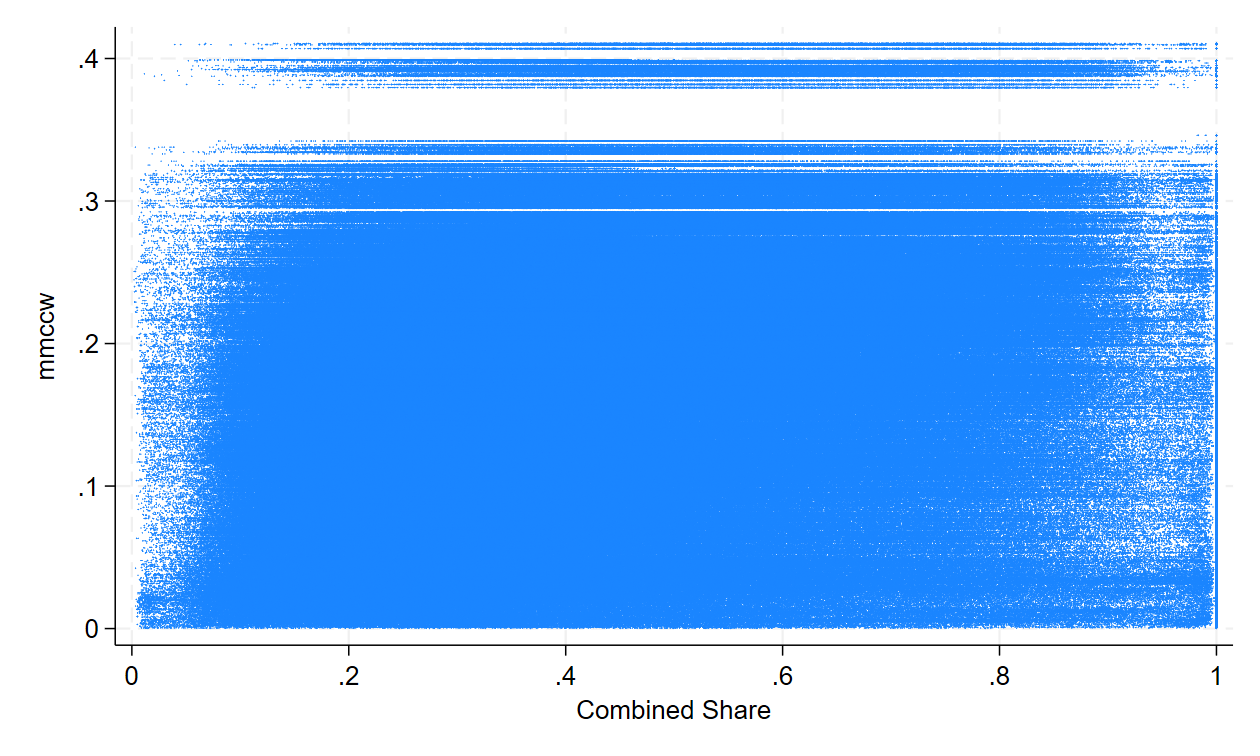}
    \caption{$MMC^{CW}$ and $CS$(0.23)}
    \end{subfigure}
    \begin{subfigure}[b]{0.32\textwidth}
    \includegraphics[width=\textwidth]{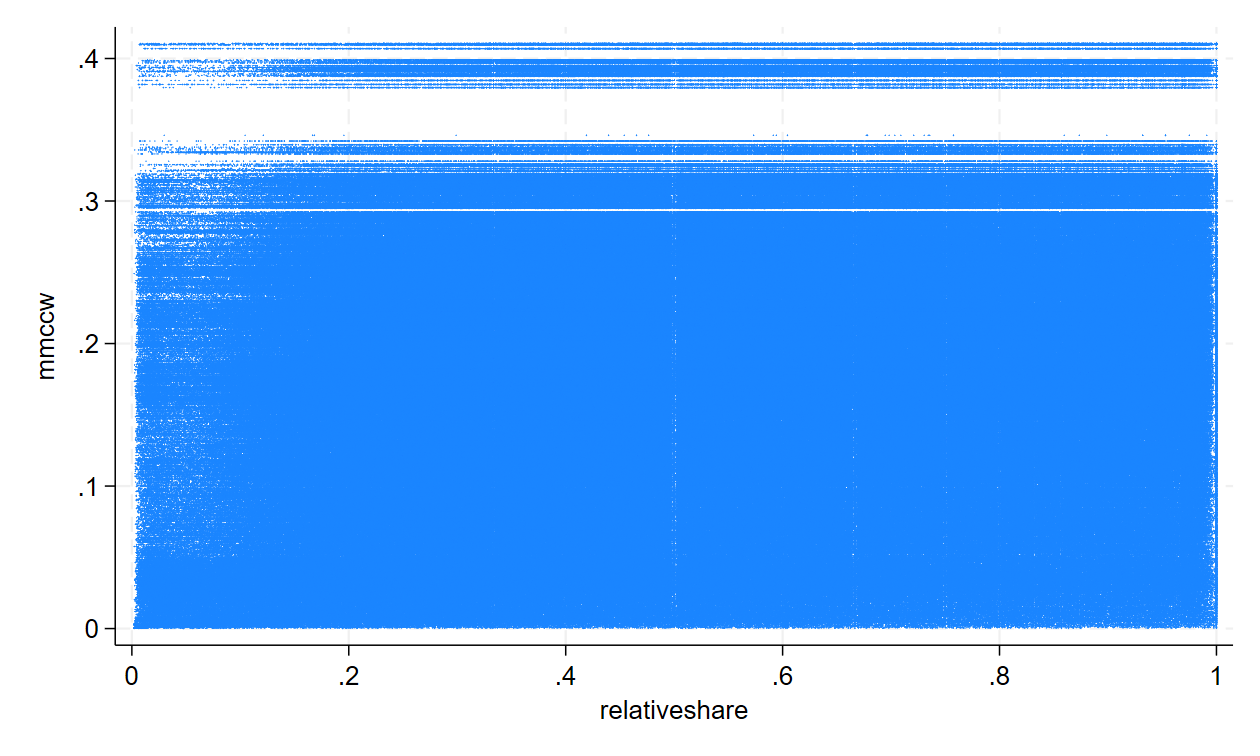}
    \caption{$MMC^{CW}$ and $RS$(0.13)}
    \end{subfigure}
    \centering
    \begin{subfigure}[b]{0.32\textwidth}
    \includegraphics[width=\textwidth]{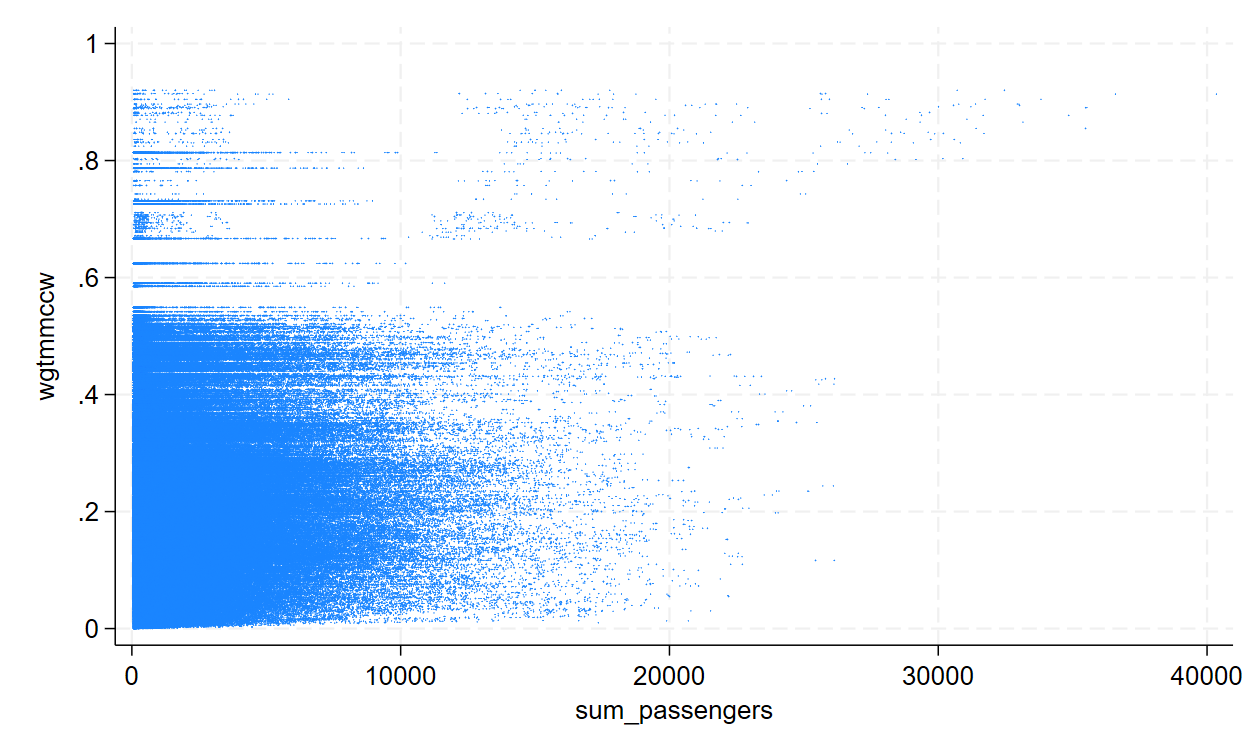}
    \caption{$MMC^{CW}_{Wgt}$ and $TP$(-0.02)}
    \end{subfigure}
    \begin{subfigure}[b]{0.32\textwidth}
    \includegraphics[width=\textwidth]{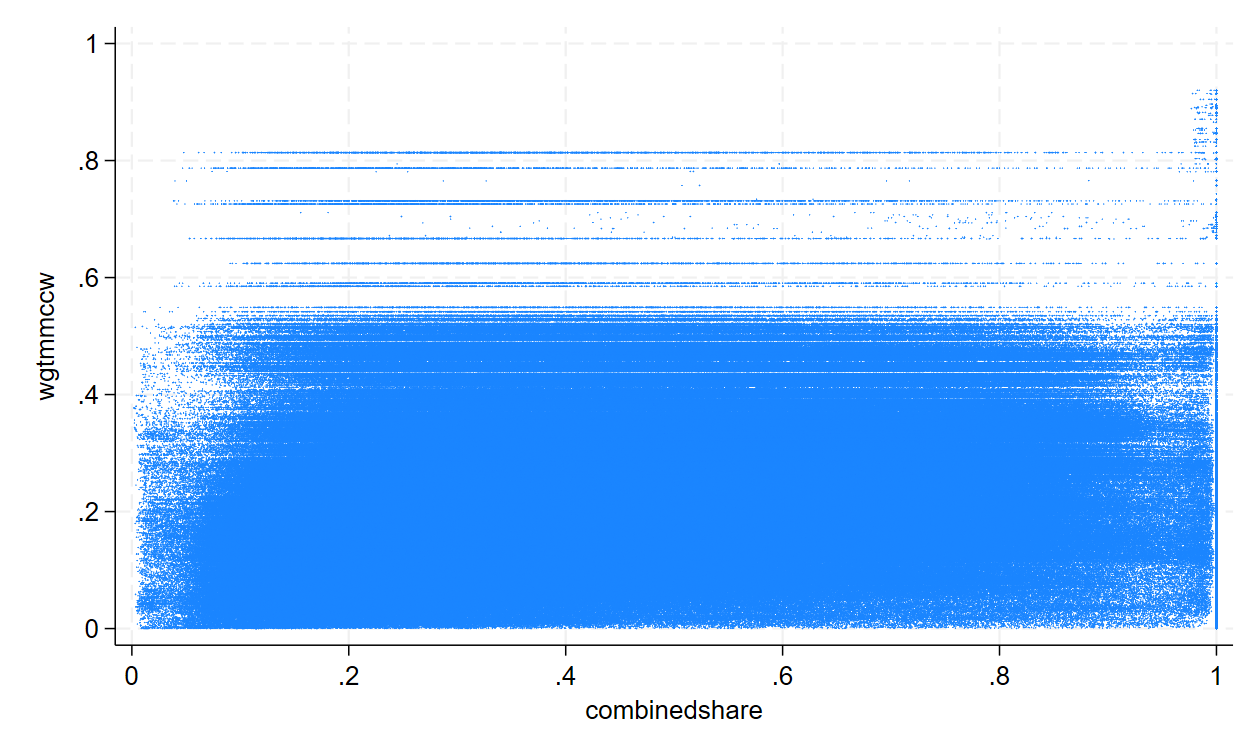}
    \caption{$MMC^{CW}_{Wgt}$ and $CS$(0.26)}
    \end{subfigure}
    \begin{subfigure}[b]{0.32\textwidth}
    \includegraphics[width=\textwidth]{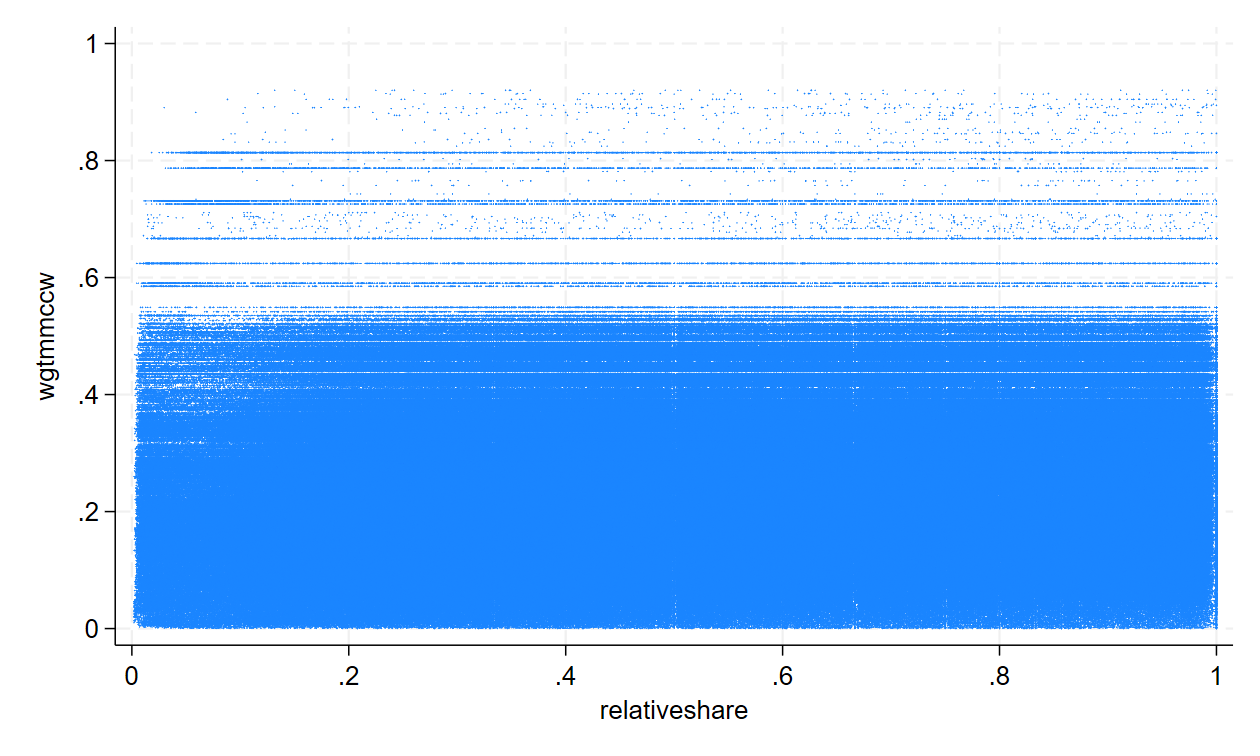}
    \caption{$MMC^{CW}_{Wgt}$ and $RS$(0.08)}
    \end{subfigure}
    Figure 4: Pairwise Scatter Plots of Independent Variables
\end{figure}

The figure above features on the scatter plots between independent variables appeared in the regression, and the correlation coefficients are reported in the bracket behind every label. It is obvious that independent variables incorporated in this paper have little correlation with each other, indicating a low risk of multicollinarity.

\end{document}